\documentclass[
 preprint,
 superscriptaddress,
 amsmath,
 amssymb,
 floatfix,
 showkeys
]{revtex4-2}
\usepackage{graphicx}
\usepackage[version=4]{mhchem}
\usepackage{hyperref}

\graphicspath{{./images/}{./fig/}}

\makeatletter
\def\Eprint#1#2{}
\makeatother

\begin{document}
\title{Nanocavity Confinement by Orthogonal Valley- and SSH- Topological Interfaces In Glide-Symmetric Photonic Crystal Structures}

\author{Takahiro Uemura}
\affiliation{Department of Physics, Institute of Science Tokyo, 2-12-1 Ookayama, Meguro-ku, Tokyo 152-8551, Japan}
\affiliation{Basic Research Laboratories, NTT Inc., 3-1 Morinosato-Wakamiya, Atsugi-shi, Kanagawa 243-0198, Japan}

\author{Wei Dai}
\affiliation{Department of Physics, Institute of Science Tokyo, 2-12-1 Ookayama, Meguro-ku, Tokyo 152-8551, Japan}
\affiliation{Department of Electrical and Electronic Engineering, The University of Tokyo, 2-11-16 Yayoi, Bunkyo-ku, Tokyo 113-0032, Japan}

\author{Yuto Moritake}
\affiliation{Department of Physics, Institute of Science Tokyo, 2-12-1 Ookayama, Meguro-ku, Tokyo 152-8551, Japan}
\affiliation{Institute of Industrial Science, The University of Tokyo, 4-6-1 Komaba, Meguro-ku, Tokyo 153-8505, Japan}

\author{Masaaki Ono}
\affiliation{Basic Research Laboratories, NTT Inc., 3-1 Morinosato-Wakamiya, Atsugi-shi, Kanagawa 243-0198, Japan}
\affiliation{Nanophotonics Center, NTT Inc., 3-1 Morinosato-Wakamiya, Atsugi-shi, Kanagawa 243-0198, Japan}

\author{Eiichi Kuramochi}
\affiliation{Basic Research Laboratories, NTT Inc., 3-1 Morinosato-Wakamiya, Atsugi-shi, Kanagawa 243-0198, Japan}
\affiliation{Nanophotonics Center, NTT Inc., 3-1 Morinosato-Wakamiya, Atsugi-shi, Kanagawa 243-0198, Japan}

\author{Masaya Notomi}
\email{notomi@phys.sci.isct.ac.jp}
\affiliation{Department of Physics, Institute of Science Tokyo, 2-12-1 Ookayama, Meguro-ku, Tokyo 152-8551, Japan}
\affiliation{Basic Research Laboratories, NTT Inc., 3-1 Morinosato-Wakamiya, Atsugi-shi, Kanagawa 243-0198, Japan}
\affiliation{Nanophotonics Center, NTT Inc., 3-1 Morinosato-Wakamiya, Atsugi-shi, Kanagawa 243-0198, Japan}

\begin{abstract}
Valley photonic crystals enable valley-dependent transport and chirality-selective emission, but incorporating wavelength-scale localization remains challenging. Existing valley-photonic-crystal cavities rely on finite defects or local lattice modifications that require structure-specific optimization and offer limited continuous control. Here, we theoretically and experimentally demonstrate two-dimensional nanocavity confinement using two orthogonal domain walls in a glide-symmetric valley photonic crystal. A valley domain wall confines the guided interface mode transversely, while an SSH-like domain wall localizes it longitudinally. Starting from a glide-symmetry-protected Dirac point in a bearded-interface waveguide, controlled displacements of adjacent triangular holes open a topological gap in the continuous guided-mode dispersion. The displacement amplitude $\Delta R$ tunes the gap, mode volume, and intrinsic radiative $Q$ factor. Implemented in a silicon photonic-crystal slab, the structure exhibits localized resonances within the topological mode gap and systematic spectral tuning with $\Delta R$. The maximum measured loaded $Q$ factor is $1.2\times10^{4}$. This approach enables continuously tunable, high-$Q$ nanocavities integrated into topological waveguide networks for compact resonant devices and enhanced light--matter interactions.
\end{abstract}

\maketitle

\section{Introduction}

Topological photonics provides a versatile framework for controlling optical propagation and localization through band topology and crystalline symmetries\cite{PhysRevLett.114.223901, doi:10.1126/science.aaq0327, doi:10.1126/sciadv.aaw4137}. In two-dimensional photonic crystals, valley photonic crystals (VPhCs) exploit the inequivalent $K$ and $K'$ valleys created by inversion-symmetry breaking, which exhibit opposite Berry curvatures and valley-contrasting phase windings \cite{Dong2017, Shalaev2019, Arora2021, Shao2020}. These properties enable valley-dependent transport and directional coupling of circularly polarized emitters \cite{He2019ValleyRouting, Mehrabad2023ChiralFilter}. Although valley-Hall waveguides are not generically immune to disorder-induced backscattering \cite{Rosiek2023Backscattering}, glide-symmetric bearded interfaces can support high transmission through engineered $120^\circ$ bends, including in the slow-light regime \cite{Yoshimi:20, Yoshimi:21, dai2023high}. Such waveguides therefore provide an attractive platform for integrated topological photonic circuits.

Introducing wavelength-scale localization into these waveguides would enable resonant filtering and enhanced light--matter interactions. In conventional W1 line-defect waveguides \cite{PhysRevLett.87.253902}, coupling between a propagating guided mode and a localized nanocavity mode is well established. Localized cavity modes can also be created by modifying the sizes or positions of nearby holes or by tapering the waveguide width to form a mode-gap cavity \cite{Song2005, 10.1063/1.2167801}. In VPhC waveguides, however, the guiding channel is an extended interface between two bulk photonic crystals, and a general strategy for introducing a localized cavity mode has not yet been established. Recent approaches based on point defects or finite-length perturbations can produce localized states \cite{Hallacy2025, Li2025ValleyDefectCavity}, but their properties typically depend on the detailed geometry of the defect region, which must be optimized for each design.

Topological domain walls provide an alternative route to nanocavity formation with more systematic control of the cavity properties. Zero-dimensional states and associated lasing have been demonstrated in resonator, waveguide, polariton, and photonic-crystal-cavity arrays \cite{MoritakeOnoNotomi+2022+2183+2189, doi:10.1126/sciadv.abf8049, PhysRevLett.120.113901, St-Jean2017, Zhao2018, Han2019}, as well as in higher-order corner states, closed topological edge channels, and crystalline defects \cite{Ota:19, Wang2025TopologicalCavities, Zhao2025SlowLightCavities,Liu2021BulkDisclination, Hwang2024, Cui2025DisclinationNanocavities}. Among these designs, Ota \textit{et al.} extended the concept of Su--Schrieffer--Heeger (SSH) model from discrete lattices \cite{PhysRevLett.42.1698, asboth2016short, Meier2016, Atala2013, Lohse2016} to the continuous guided bands of one-dimensional photonic-crystal nanobeams, realizing a zero-dimensional interface state and single-mode nanocavity lasing between nanobeams with distinct Zak phases \cite{Ota2018}. However, extending this approach to a two-dimensional photonic-crystal waveguide remains challenging because transverse guiding and longitudinal topological localization must be implemented within the same continuous interface band. Shifted two-dimensional photonic-crystal interfaces with opposite Dirac masses have been predicted to open a guided-mode gap and support a domain-wall state \cite{10.1063/5.0186703}, but such a design has not yet been implemented in a silicon photonic-crystal slab suitable for optical nanocavity confinement.

Here, we combine two topological mechanisms acting at different dimensional levels and in orthogonal spatial directions. A transverse valley-Hall domain wall between two-dimensional VPhC bulk domains forms the waveguide and confines the guided interface mode across the waveguide. Within this valley interface, a longitudinal SSH-like domain wall between waveguide sections with distinct Zak phases localizes the mode along the propagation direction. The intersection of these two domain walls therefore produces a zero-dimensional cavity confined in both in-plane directions, exploiting a geometric and topological degree of freedom unavailable in conventional one-dimensional SSH systems. Starting from a glide-symmetry-protected Dirac point\cite{Plotnik2014, Yoshimi:20, PhysRevB.94.195109, PhysRevB.106.064304, Mock:20, PhysRevA.111.033513}, controlled displacements of triangular holes adjacent to the interface open a topological gap in the guided-mode dispersion. Reversing the displacement produces two waveguide sections with distinct Zak phases, and their domain wall supports a localized cavity mode. The displacement amplitude $\Delta R$ provides continuous control of the confinement, in contrast to approaches based on finite defects whose dimensions change in discrete lattice increments. We implement this concept in a silicon photonic-crystal slab and experimentally observe localized resonances inside the induced mode gap, with a maximum measured loaded $Q$ factor of $1.2\times10^4$. This approach establish a simple route to continuously tunable nanocavities embedded directly within topological waveguides and provide a platform for combining localized resonances with valley-dependent photonic transport.

\section{Symmetry-based SSH localization in a glide-symmetric interface band}

The symmetry argument developed below starts from the full-vector Maxwell eigenproblem and is therefore not restricted to structures that are invariant along the $z$ direction. To isolate the band topology from vertical radiation and slab-mode hybridization, however, the numerical examples in this section use a two-dimensional effective-index model that is invariant along $z$. Unless otherwise noted, the calculations were performed using the finite-element method (FEM) implemented in COMSOL Multiphysics.

\subsection{Opening an SSH-like gap in a continuous interface band}

\begin{figure}[htbp]
	\centering
	\includegraphics[width=1\linewidth]{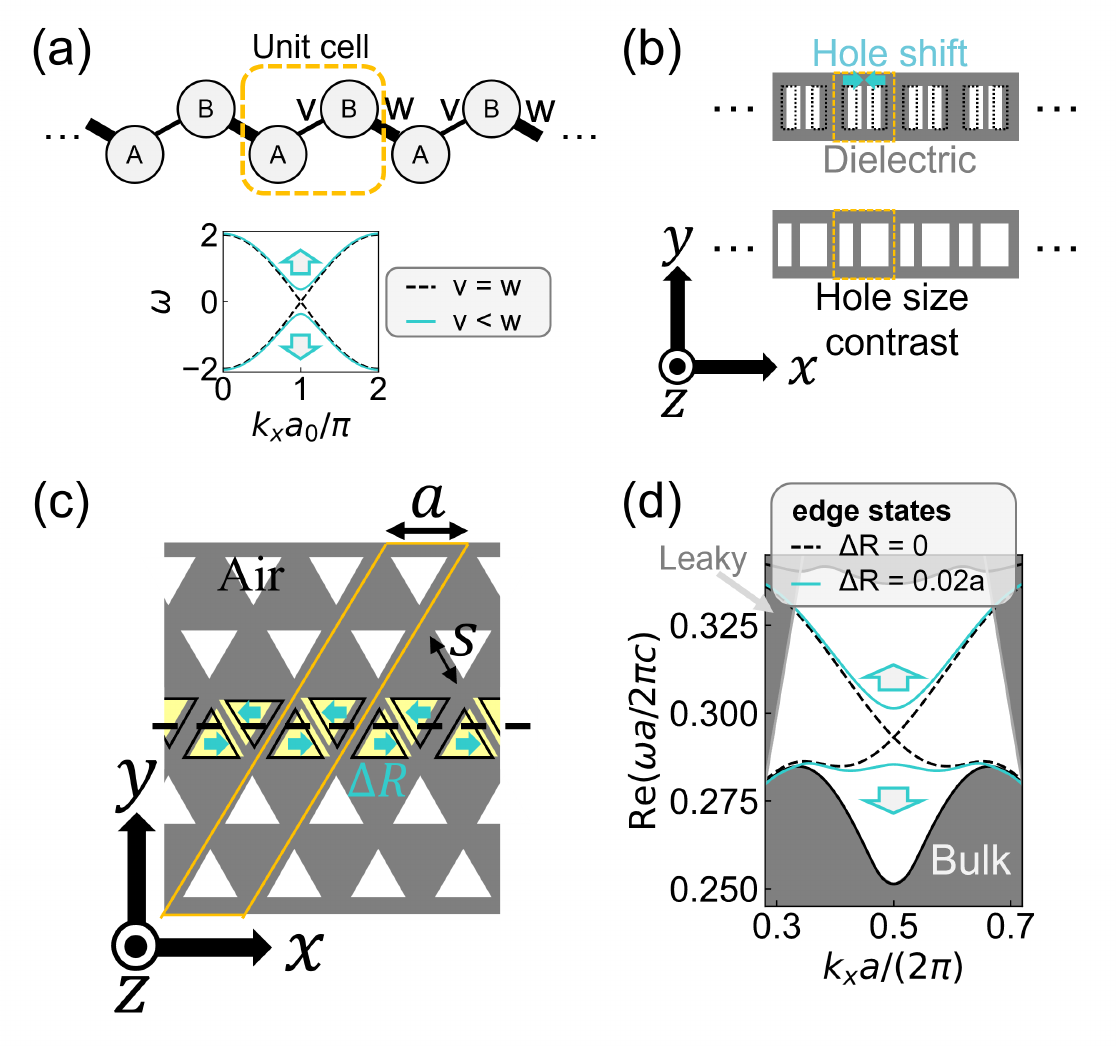}
	\caption{
        (a) Schematic of the Su--Schrieffer--Heeger (SSH) model, with intracell and intercell couplings $v$ and $w$, respectively.
        (b) Representative one-dimensional photonic-crystal nanobeam realizations, in which geometric modulation produces an effective SSH dimerization along an isolated nanobeam \cite{Ota2018}.
        (c) Glide-symmetric interface between two VPhCs. The dashed outline indicates the unit cell, and the arrows indicate the opposite displacements $\pm\Delta R$ of the triangular holes adjacent to the interface.
        (d) Dispersion of the guided interface band. The unperturbed glide-symmetric structure ($\Delta R=0$, black dashed lines) has a Dirac point at the Brillouin-zone boundary, whereas the perturbation ($\Delta R=0.02a$, green solid lines) opens a mode gap directly in the continuous interface-band dispersion.
        }
	\label{fig:fig1}
\end{figure}
Figure~\ref{fig:fig1}(a) summarizes the essential idea of the Su--Schrieffer--Heeger (SSH) model, in which a one-dimensional lattice is composed of alternating intracell and intercell couplings with amplitudes \(v\) and \(w\), respectively. The Bloch Hamiltonian of the SSH model can be written as
%
\begin{align}
 \label{eq:ssh_hamiltonian}
 H_\mathrm{SSH} &=
 \left[
  \begin{matrix}
   0 & v+ we^{-ik_x a} \\
   v + we^{ik_x a} & 0
  \end{matrix}
  \right]
\end{align}
where $v$ and $w$ are the intracell and intercell hopping amplitudes, respectively, and $a$ is the lattice constant. When the two couplings are equal ($v=w$), the primitive cell can be reduced to a single site; folding the corresponding band into the doubled Brillouin zone produces a linear crossing at $k_x=\pi/a$. Dimerization ($v\neq w$) opens a bandgap at this crossing.
For the unit-cell convention used in Eq.~\eqref{eq:ssh_hamiltonian}, the lower band has winding number $0$ and Zak phase $0$ when $v>w$, and winding number $1$ and Zak phase $\pi$ when $v<w$. The gap closes at $v=w$, marking the transition between the two phases \cite{PhysRevLett.42.1698,asboth2016short}.

Figure~\ref{fig:fig1}(b) shows how this SSH mechanism can be implemented in photonic-crystal nanobeam cavities. Periodic geometric modulation---for example, shifts in the air-hole positions or alternating hole sizes---emulates the two hopping amplitudes in the optical domain. The modulation produces an effective dimerization of the guided mode and opens a controllable photonic bandgap, in direct analogy with the tight-binding SSH chain \cite{Ota2018}.
Extending the SSH concept from a discrete chain to a one-dimensional nanobeam is relatively natural because both systems are described along a single spatial direction. Extending it to a two-dimensional photonic-crystal waveguide is less direct because the relevant topology must be introduced within a continuous interface band. In our design, glide symmetry identifies both the degeneracy and the symmetry-allowed perturbation required to open an SSH-like gap. Figure~\ref{fig:fig1}(c) illustrates the resulting two-dimensional glide-symmetric photonic-crystal waveguide. The structure consists of a triangular lattice with lattice constant $a=470\,\mathrm{nm}$ and triangular holes with side length $0.8a$. The geometry is modeled using a two-dimensional effective-index approximation for transverse-electric (TE)-polarized modes, with $n_\mathrm{eff}=2.6$ to represent the silicon slab.
The triangular-lattice valley photonic-crystal platform is advantageous because it supports a relatively large bulk bandgap, allowing the glide-symmetric interface band and the SSH-like mode gap to be treated within a well-isolated spectral window.
The yellow dashed outline in Fig.~\ref{fig:fig1}(c) denotes the unit cell. The black dashed triangles indicate the hole positions in the unperturbed structure. This structure is invariant under a glide reflection about the dashed horizontal line. The glide symmetry enforces a degeneracy at the Brillouin-zone boundary, $k_xa=\pi$ \cite{PhysRevB.103.235110, PhysRevB.94.195109, PhysRevB.106.064304, PhysRevA.111.033513, 10.1063/5.0186703}, as indicated by the black dashed curves in Fig.~\ref{fig:fig1}(d).
To open a topological mode gap, we displace the triangular holes adjacent to the photonic-crystal interface by $+\Delta R$ and $-\Delta R$ along $x$ on opposite sides of the glide plane. The yellow triangular markers in Fig.~\ref{fig:fig1}(c) indicate the displaced positions. This perturbation lifts the Dirac-point degeneracy and opens a bandgap, analogously to the periodic modulation of the nanobeams in Fig.~\ref{fig:fig1}(b). As shown below, the perturbation preserves the symmetry required for an SSH-like topology. The same strategy can in principle be applied to other two-dimensional lattices, provided that the unperturbed waveguide supports an appropriate glide-protected degeneracy and that the perturbation satisfies the required symmetry. Because the original Dirac point lies below the light line, the corresponding modes of the periodically perturbed waveguide remain nonradiative in this frequency range. This is an important advantage of the present platform: opening an SSH-like gap requires a doubled period along the waveguide, and band folding in conventional photonic-crystal waveguides can move the relevant modes above the light line.

\begin{figure}[htbp]
	\centering
	\includegraphics[width=0.85\linewidth]{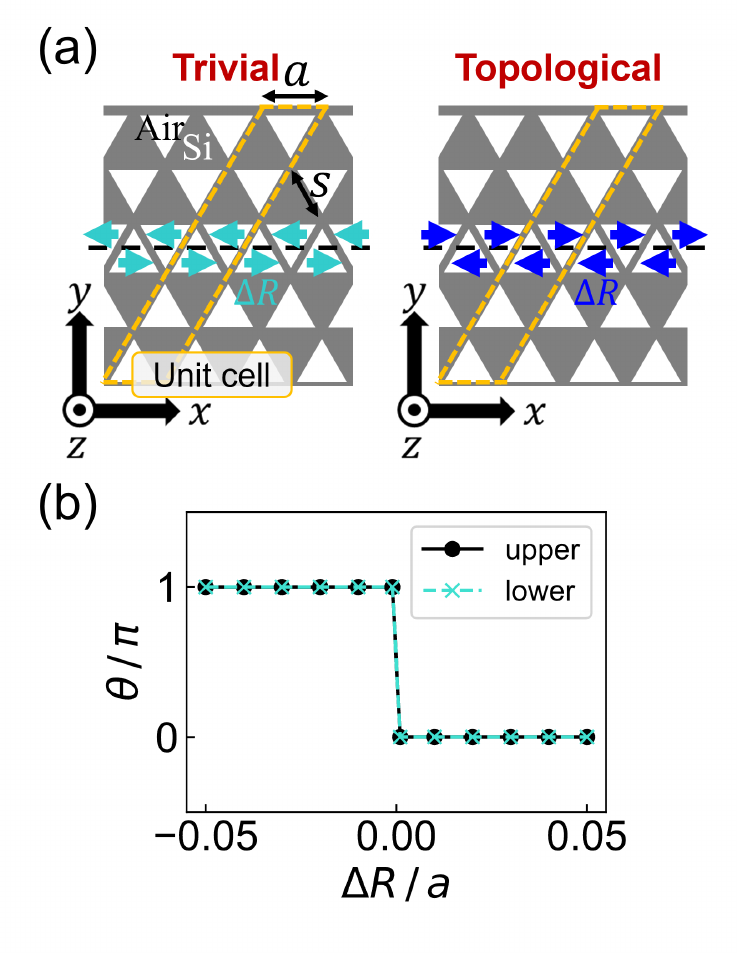}
	\caption{
		(a) Geometric structure of the photonic crystal for $\Delta R > 0$ (left) and $\Delta R < 0$ (right).
		(b) Zak phases of the upper and lower interface bands calculated from the cell-periodic six-component Bloch eigenstates. The results are shown for $\Delta R$ values at intervals of $a/100$, with additional points at $\Delta R = \pm a/1000$. At $\Delta R=0$, the two bands touch and their individual Zak phases are undefined.
		}
	\label{fig:fig2}
\end{figure}
The topology of the gapped interface bands can be characterized by the Zak phase \cite{PhysRevLett.62.2747}. We evaluated it using six-component electromagnetic eigenstates obtained from a two-dimensional FEM model with Bloch-periodic boundary conditions. The physical Bloch fields are written as
\begin{align}
 \boldsymbol{\mathcal{E}}_{n\boldsymbol{k}}(\boldsymbol{r})
 &=e^{i\boldsymbol{k}\cdot\boldsymbol{r}}
 \boldsymbol{E}_{n\boldsymbol{k}}(\boldsymbol{r}),\\
 \boldsymbol{\mathcal{H}}_{n\boldsymbol{k}}(\boldsymbol{r})
 &=e^{i\boldsymbol{k}\cdot\boldsymbol{r}}
 \boldsymbol{H}_{n\boldsymbol{k}}(\boldsymbol{r}),
\end{align}
where $\boldsymbol{E}_{n\boldsymbol{k}}$ and $\boldsymbol{H}_{n\boldsymbol{k}}$ are the cell-periodic field envelopes obtained after factoring out the Bloch phase. These envelopes were combined into a normalized six-component state, and the Wilson loop was constructed from energy-weighted overlaps between neighboring wavevectors. Because the overlaps are evaluated between the cell-periodic Bloch states, the Bloch phase factor $e^{i\boldsymbol{k}\cdot\boldsymbol{r}}$ is not included in the overlap integral. The numerical procedure is detailed in \ref{sec:zak_phase}, and the discretized Zak phase is given by Eq.~\eqref{eq:zak_phase_numerical}.

We sampled the Brillouin zone from $k_x=0$ to $2\pi/a$ using 40 equal intervals, $\Delta\boldsymbol{k}=[2\pi/(40a),0,0]^\top$. Figure~\ref{fig:fig2}(b) shows that, for the chosen unit cell, both the lower and upper interface bands have Zak phase $\pi$ for $\Delta R<0$ and $0$ for $\Delta R>0$. The two bands become degenerate at $\Delta R=0$, where their individual Zak phases are not defined. Thus, reversing the displacement changes the Zak phase of each gapped interface band by $\pi$, consistent with an SSH-like topological transition in the continuous guided-mode dispersion.

\subsection{Symmetry criterion for the SSH-like topology}

The Zak-phase calculation establishes the phase inversion for the two-dimensional model. We now identify the symmetry condition responsible for the SSH-like form of the full-vector effective Hamiltonian.
\begin{figure}[htbp]
	\centering
	\includegraphics[width=1\linewidth]{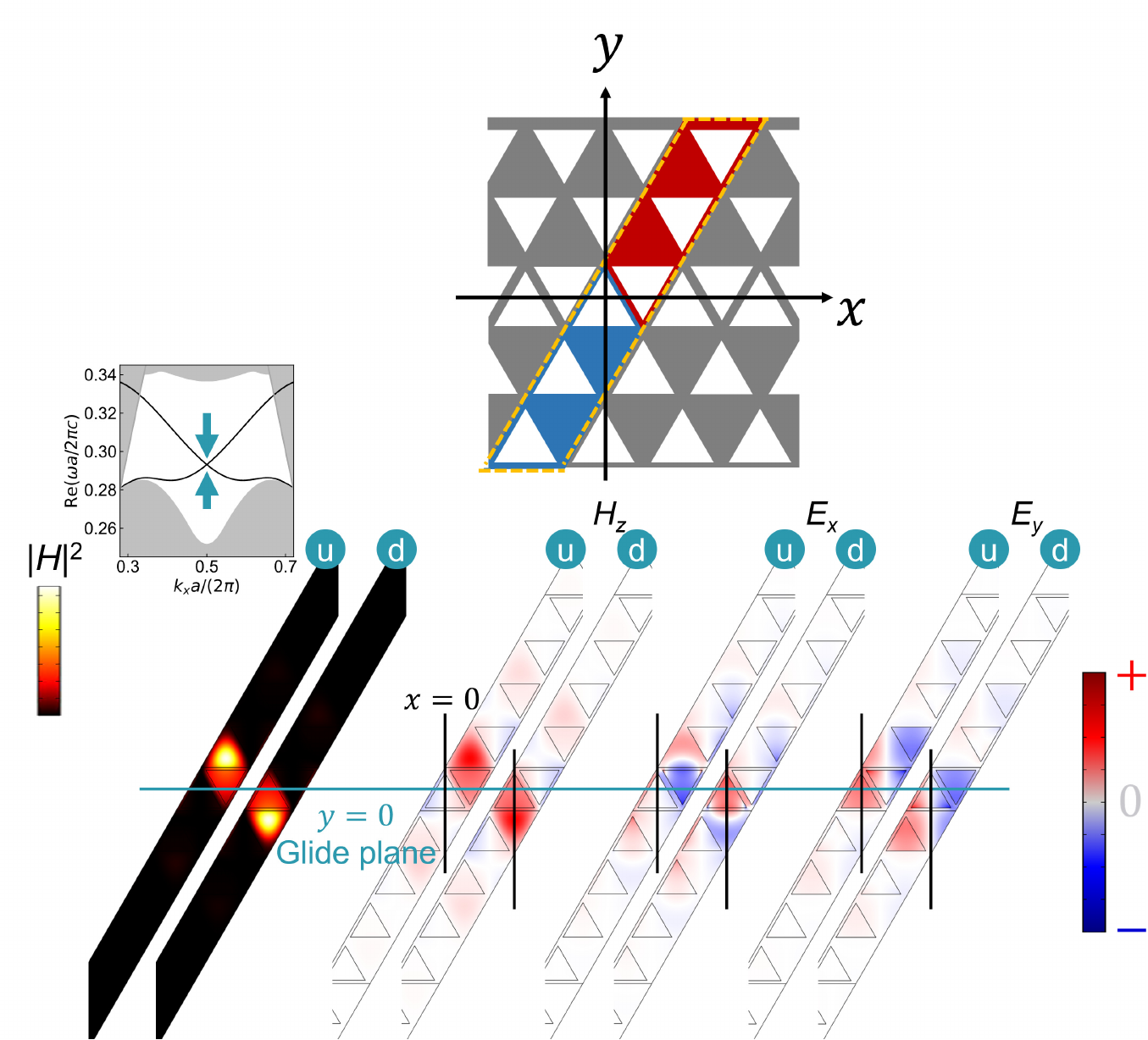}
	\caption{
		 Distributions of ${|\boldsymbol{H}|}^2$, $H_z$, $E_x$, and $E_y$ for the basis states at the Dirac point.
		The labels $\mathrm{u}$ and $\mathrm{d}$ denote $\boldsymbol{\varphi}_{\mathrm{u}k_0}^{(0)}$ and $\boldsymbol{\varphi}_{\mathrm{d}k_0}^{(0)}$, respectively.
		}
	\label{fig:fig3}
\end{figure}
To identify the relevant symmetry constraints, we consider the effective Hamiltonian of the photonic-crystal waveguide near the Dirac point at $k_x=\pi/a$. We use the two degenerate unperturbed electromagnetic modes at $k_0=\pi/a$ as a basis [Fig.~\ref{fig:fig3}], denoting their magnetic and electric components by $\boldsymbol{\varphi}^{(0)}_{\alpha k_0}$ and $\boldsymbol{\psi}^{(0)}_{\alpha k_0}$, respectively. These states are standing waves with opposite mirror parities of $H_z$ with respect to the $x$ axis. This basis allows the dispersion near the Dirac point to be approximated from the eigenmode profiles at the degeneracy.

Applying $\boldsymbol{k}\cdot\boldsymbol{p}$ perturbation theory (\ref{sec:kp_method}) gives the following effective Hamiltonian near the Dirac point:

\begin{align}
	\label{eq:effective_hamiltonian}
  H' =
  \begin{bmatrix}
   m_\mathrm{D} & \kappa + i v_\mathrm{D} \delta k_x \\
   \kappa - i v_\mathrm{D} \delta k_x & - m_\mathrm{D}
  \end{bmatrix}
\end{align}
where \(\delta k_x = k_x - \pi/a\) is the deviation of the wavevector from the Dirac point. The dielectric coupling coefficient \(\kappa\) and the Dirac velocity parameter \(v_\mathrm{D}\) are defined by
\begin{align}
 \label{eq:kappa_vD_def}
 \Delta\eta(\boldsymbol{r})
 &:=\frac{1}{\varepsilon(\boldsymbol{r})+\Delta\varepsilon(\boldsymbol{r})}
   -\frac{1}{\varepsilon(\boldsymbol{r})}, \nonumber\\
 \kappa
 &:=M_{\mathrm{u}\mathrm{d}}
 =\frac{\omega_D^2}{c^2}
 \int_{\mathrm{u.c.}}\mathrm{d}^3r\,
 \varepsilon^2(\boldsymbol{r})\Delta\eta(\boldsymbol{r})
 \boldsymbol{\psi}_{\mathrm{u}k_0}^{(0)*}(\boldsymbol{r})\cdot
 \boldsymbol{\psi}_{\mathrm{d}k_0}^{(0)}(\boldsymbol{r}) \nonumber\\
 &\simeq
 -\frac{\omega_D^2}{c^2}
 \int_{\mathrm{u.c.}}\mathrm{d}^3r\,
 \Delta\varepsilon(\boldsymbol{r})
 \boldsymbol{\psi}_{\mathrm{u}k_0}^{(0)*}(\boldsymbol{r})\cdot
 \boldsymbol{\psi}_{\mathrm{d}k_0}^{(0)}(\boldsymbol{r}), \nonumber\\
 v_\mathrm{D}
 &:= -iP_{\mathrm{u}\mathrm{d}}.
\end{align}
Here, \(\Delta\eta\) is the exact inverse-permittivity change. The approximate expression follows from
\(\varepsilon^2\Delta\eta=-\Delta\varepsilon+\mathcal{O}(\Delta\varepsilon^2)\).
The matrix element \(P_{\alpha\beta}\) arises from the \(\boldsymbol{k}\cdot\boldsymbol{p}\) expansion and is given in \ref{sec:kp_method}. The diagonal detuning term is
\begin{align}
 \label{eq:delta_g}
 m_\mathrm{D}
 &:=\frac{M_{\mathrm{u}\mathrm{u}}-M_{\mathrm{d}\mathrm{d}}}{2} \nonumber\\
 &=\frac{\omega_D^2}{2c^2}
 \int_{\mathrm{u.c.}}\mathrm{d}^3r\,
 \varepsilon^2(\boldsymbol{r})\Delta\eta(\boldsymbol{r})
 \left(
  \left|\boldsymbol{\psi}_{\mathrm{u}k_0}^{(0)}\right|^2
  -\left|\boldsymbol{\psi}_{\mathrm{d}k_0}^{(0)}\right|^2
 \right) \nonumber\\
 &\simeq
 \frac{\omega_D^2}{2c^2}
 \int_{\mathrm{u.c.}}\mathrm{d}^3r\,
 \Delta\varepsilon(\boldsymbol{r})
 \left(
  \left|\boldsymbol{\psi}_{\mathrm{d}k_0}^{(0)}\right|^2
  -\left|\boldsymbol{\psi}_{\mathrm{u}k_0}^{(0)}\right|^2
 \right).
\end{align}
Because Eq.~\eqref{eq:delta_g} is written in terms of three-dimensional vector fields and a volume integral, the symmetry criterion itself is applicable to both two-dimensional structures and finite-thickness photonic-crystal slabs. The two-dimensional calculation in this section is used only to provide a transparent numerical illustration of the phase transition.
To compare Eq.~\eqref{eq:effective_hamiltonian} with the SSH model, we expand Eq.~\eqref{eq:ssh_hamiltonian} about $k_x=\pi/a$ by defining $\delta k_x=k_x-\pi/a$. To first order,
\begin{align}
 \label{eq:ssh_hamiltonian_expanded}
 H_\mathrm{SSH} \approx
 \left[
 \begin{matrix}
  0 & (v - w) + iwa\, \delta k_x \\
  (v - w) - iwa\,\delta k_x & 0
 \end{matrix}
 \right]
\end{align}

To interpret the condition \(m_\mathrm{D} = 0\) in terms of symmetry operations, we define the glide operator \(\hat{G}\) and the mirror operator \(\hat{M}_x\) as
\begin{align}
	\label{eq:operators}
  \hat{G} = \left\{ m_y \middle| \frac{a}{2} \hat{\boldsymbol{x}} \right\}, \quad
  \hat{M}_x = \left\{ m_x \middle| 0 \right\}
\end{align}
Here, $m_y$ and $m_x$ denote the reflections $y\mapsto-y$ and $x\mapsto-x$, respectively. As shown in Fig.~\ref{fig:fig3}, the intensities of the basis states $\boldsymbol{\psi}_{\mathrm{u}k_0}^{(0)}$ and $\boldsymbol{\psi}_{\mathrm{d}k_0}^{(0)}$ are exchanged by $\hat{G}$ and are individually invariant under $\hat{M}_x$. Equation~\eqref{eq:delta_g} therefore gives $m_\mathrm{D}=0$ when the scalar perturbation weight $\varepsilon^2\Delta\eta$ is invariant under the combined glide--mirror operation. In the weak-permittivity-perturbation form, this reduces to the following condition on $\Delta\varepsilon(\boldsymbol{r})$:
\begin{align}
 \left( \hat{G} \hat{M}_x \right) \Delta \varepsilon \left( \boldsymbol{r} \right) = \Delta \varepsilon \left( -x + \frac{a}{2}, -y, z \right) = \Delta \varepsilon \left( \boldsymbol{r} \right)
\end{align}
Under this condition, the effective Hamiltonian maps, to leading order near the Dirac point, onto the SSH form. The Zak phase is a global quantity that requires information over the entire Brillouin zone, as evaluated in Fig.~\ref{fig:fig2}(b), whereas this local mapping identifies the symmetry condition required for the SSH-like gap. The proposed symmetric hole displacement preserves $\hat{G}\hat{M}_x$ and therefore eliminates the diagonal detuning term to leading order.

\subsection{Domain-wall localization in the continuous interface band}

\begin{figure}[htbp]
	\centering
	\includegraphics[width=\linewidth,height=0.70\textheight,keepaspectratio]{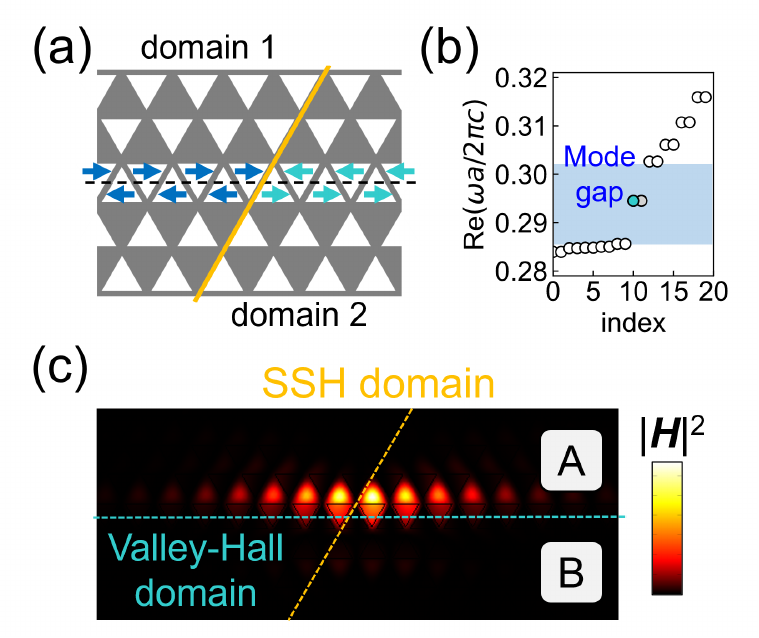}
	\caption{
		(a) Domain-wall structure of the photonic-crystal waveguide. The left and right domains have $\Delta R=-0.02a$ and $+0.02a$, respectively.
		(b) Eigenfrequencies of the waveguide supercell. The blue shaded region denotes the mode gap. Green and gray markers indicate the in-gap modes for the two inversion-related domain-wall configurations.
		(c) Distribution of $|\boldsymbol{H}|^2$ for the nanocavity mode of the configuration with $\Delta R=-0.02a$ in the left SSH domain and $+0.02a$ in the right SSH domain. The longitudinal SSH-like domain wall intersects the transverse valley-Hall domain wall, which coincides with the glide plane. The upper and lower valley-Hall domains are labeled A and B, respectively.
		}
	\label{fig:fig4}
\end{figure}
Boundary states at interfaces between distinct topological phases are a central feature of topological systems. Because air does not provide a common spectral gap for the guided band considered here, a topological state is not generally expected at the outer boundary of the photonic crystal. We therefore analyze a longitudinal SSH-like domain wall embedded in the transverse valley-Hall domain wall of the photonic-crystal waveguide [Fig.~\ref{fig:fig4}(a)]. The valley-Hall domain wall, which coincides with the glide plane of the unperturbed bearded interface, separates the upper and lower VPhC domains and confines the guided interface mode in the transverse direction. Along this interface, the left and right SSH-like domains have $\Delta R=-0.02a$ and $+0.02a$, respectively, and their lower interface bands have Zak phases $\pi$ and $0$ for the chosen unit cell. The periodic supercell contains two SSH-like domain walls and therefore supports two in-gap domain-wall states.

Figure~\ref{fig:fig4}(b) shows two eigenfrequencies inside the mode gap. The two states arise because the periodic supercell contains two SSH-like domain walls. The $|\boldsymbol{H}|^2$ profile shown in Fig.~\ref{fig:fig4}(c) is strongly localized near the intersection of the longitudinal SSH-like domain wall and the transverse valley-Hall domain wall. The former provides longitudinal localization, while the latter provides transverse waveguide confinement, resulting in a nanocavity mode confined in both in-plane directions.
Unlike the zero mode of an ideal chiral-symmetric SSH chain, the present domain-wall modes are not necessarily pinned to the center of the gap. The local SSH-domain-wall geometry can introduce chiral-symmetry-breaking corrections and shift the boundary-mode frequency \cite{PhysRevLett.106.106802}. Nevertheless, the two SSH-domain-wall configurations obtained by reversing the signs of $\Delta R$ in the left and right domains are related by spatial inversion [green and gray markers in Fig.~\ref{fig:fig4}(b)] and therefore have the same resonance frequency. Their field profiles are localized predominantly in opposite valley-Hall domains on either side of the glide plane.
The magnetic-field distribution in Fig.~\ref{fig:fig4}(c) is concentrated in the upper valley-Hall domain, labeled A. This behavior is analogous to an SSH edge state localized on one sublattice: here, the upper and lower valley-Hall domains, A and B, play the roles of the two effective sublattices. Exchanging the left and right SSH domains transfers the mode to the lower valley-Hall domain B while preserving its resonance frequency and field extent. The intersection of the SSH-like and valley-Hall domain walls can therefore serve as a nanocavity embedded in the glide-symmetric photonic-crystal waveguide.

\section{Implementation in a silicon photonic-crystal slab and experimental observation}

We now apply the same symmetry-guided perturbation to a finite-thickness silicon photonic-crystal slab. The two-dimensional calculations in Section~2 establish the phase inversion and domain-wall localization, while the full-vector three-dimensional simulations in this section test whether the glide-protected crossing, perturbation-induced mode gap, and localized resonance persist in a realistic slab geometry.

\subsection{Design of the topological nanocavity in a silicon photonic-crystal slab}
\begin{figure}[htbp]
	\centering
	\includegraphics[width=\linewidth,height=0.68\textheight,keepaspectratio]{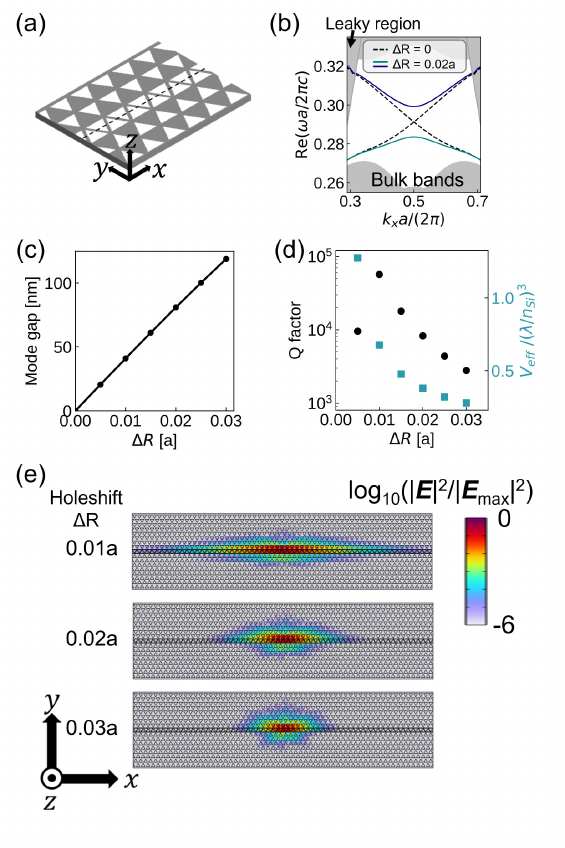}
	\caption{
  (a) Schematic of the slab-type silicon photonic-crystal waveguide with a triangular lattice. The lattice constant is $a = 440$ nm, the side length of each triangular hole is $s = 0.78a$, and the slab thickness is $h = 220$ nm.
  (b) Band dispersion of the guided interface modes along $k_x$. The glide-protected Dirac-like degeneracy appears at $k_x=\pi/a$, and a mode gap opens when the triangular-hole displacement $\Delta R$ is introduced.
  (c) Mode-gap size at the X point as a function of $\Delta R$.
  (d) Quality factor $Q$ and normalized mode volume $V_\mathrm{eff} / {(\lambda / n_\mathrm{Si})}^3$ of the domain-wall modes as functions of $\Delta R$.
  (e) Normalized $\log_{10} |\boldsymbol{E}|^2$ distributions of the domain-wall mode at $\Delta R = 0.01a$, $0.02a$, and $0.03a$.
		}
	\label{fig:fig5}
\end{figure}
Figure~\ref{fig:fig5}(a) shows the triangular-lattice silicon photonic-crystal slab waveguide. The lattice constant, triangular-hole side length, and slab thickness are $a=440\,\mathrm{nm}$, $s=0.78a$, and $h=220\,\mathrm{nm}$, respectively. The silicon refractive index is set to $n_\mathrm{Si}=3.48$. Figure~\ref{fig:fig5}(b) shows the guided-interface-mode dispersion. A Dirac-like crossing occurs at $k_x=\pi/a$, and a triangular-hole displacement $\Delta R$ lifts the degeneracy and opens a mode gap.
Figure~\ref{fig:fig5}(c) shows the wavelength-domain mode-gap width at the X point as a function of $\Delta R$. The gap increases monotonically with the displacement and reaches approximately $120\,\mathrm{nm}$ at $\Delta R=0.03a$.
Figure~\ref{fig:fig5}(d) shows the calculated $Q$ factor and normalized mode volume of the domain-wall mode as functions of $\Delta R$. The mode volume is defined as
\begin{equation}
V_\mathrm{eff}=\frac{\displaystyle\int \epsilon(\boldsymbol{r})|\boldsymbol{E}(\boldsymbol{r})|^2\,\mathrm{d}^3r}{\displaystyle\max_{\boldsymbol{r}}\left[\epsilon(\boldsymbol{r})|\boldsymbol{E}(\boldsymbol{r})|^2\right]}.
\end{equation}
Although part of the field is concentrated in the air holes, we normalize $V_\mathrm{eff}$ by the silicon wavelength scale $(\lambda/n_\mathrm{Si})^3$, which gives a conservative (larger) normalized mode volume. Increasing $\Delta R$ widens the gap and strengthens longitudinal confinement, thereby reducing the mode volume. At the smallest displacement, $\Delta R=0.005a$, the weak localization leads to appreciable in-plane leakage through the scattering boundaries of the finite simulation domain and lowers the calculated $Q$. For larger $\Delta R$, stronger localization increases out-of-plane radiation and reduces the radiative $Q$ factor.
At $\Delta R=0.01a=4.4\,\mathrm{nm}$, the calculated $Q$ factor is $5.7\times10^4$. The figure of merit $Q/[V_\mathrm{eff}/(\lambda/n_\mathrm{Si})^3]$ is also maximal at this displacement, with a value of $8.4\times10^4$. 
This trade-off between $Q$ factor and mode volume may be further optimized by modifying the transverse lattice geometry. For example, a design with weaker transverse confinement could reduce the high-spatial-frequency components responsible for out-of-plane radiation; this possibility requires separate numerical optimization.
Figure~\ref{fig:fig5}(e) shows the normalized $\log_{10}|\boldsymbol{E}|^2$ distributions for $\Delta R=0.01a$, $0.02a$, and $0.03a$. The mode is localized at the longitudinal domain wall, and its spatial extent along $x$ decreases as $\Delta R$ increases. Thus, the localization length can be systematically controlled through the triangular-hole displacement.

\subsection{Device fabrication and optical characterization}

\begin{figure}[htbp]
	\centering
	\includegraphics[width=1\linewidth]{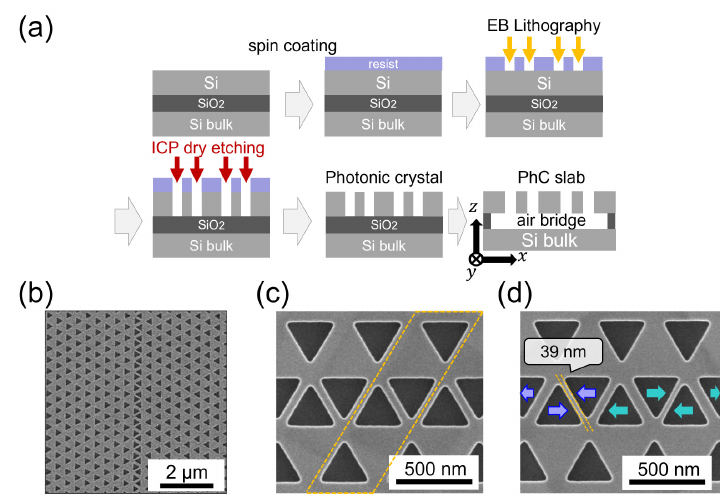}
	\caption{
			(a) Fabrication process for the photonic-crystal waveguide on a silicon-on-insulator substrate. Electron-beam lithography and inductively coupled plasma (ICP) dry etching are followed by removal of the buried \ce{SiO2} layer to form an air bridge.
  (b) SEM image of the fabricated photonic-crystal waveguide. The glide plane is located at the horizontal center of the image. The black and gray regions represent the air holes and silicon, respectively.
  (c,d) Magnified SEM images of the shifted-hole region for (c) $\Delta R=0$ and (d) $\Delta R=0.03a$.
		}
	\label{fig:fig6}
\end{figure}
The photonic-crystal waveguide was fabricated in the silicon device layer of a silicon-on-insulator (SOI) substrate using electron-beam lithography and inductively coupled plasma dry etching [Fig.~\ref{fig:fig6}(a)]. After patterning and silicon etching, the buried silicon dioxide layer was removed to form an air bridge. The silicon slab thickness and triangular-lattice constant were $220\,\mathrm{nm}$ and $440\,\mathrm{nm}$, respectively. The triangular holes had side length $0.78a$, and the displacement was varied over $\Delta R=0$, $0.005a$, $0.01a$, $0.015a$, $0.02a$, $0.025a$, and $0.03a$. Figures~\ref{fig:fig6}(b)--(d) show scanning electron microscopy (SEM) images of the fabricated structures. The displaced holes closely follow the design, and the displacement is clearly visible in Fig.~\ref{fig:fig6}(d). Even at $\Delta R=0.03a$, for which the minimum silicon bridge between adjacent holes is only $39\,\mathrm{nm}$, the structure remains well defined.

\subsection{Experimental observation of the mode-gap opening and domain-wall resonance}

\begin{figure*}[htbp]
	\centering
	\includegraphics[width=0.9\linewidth]{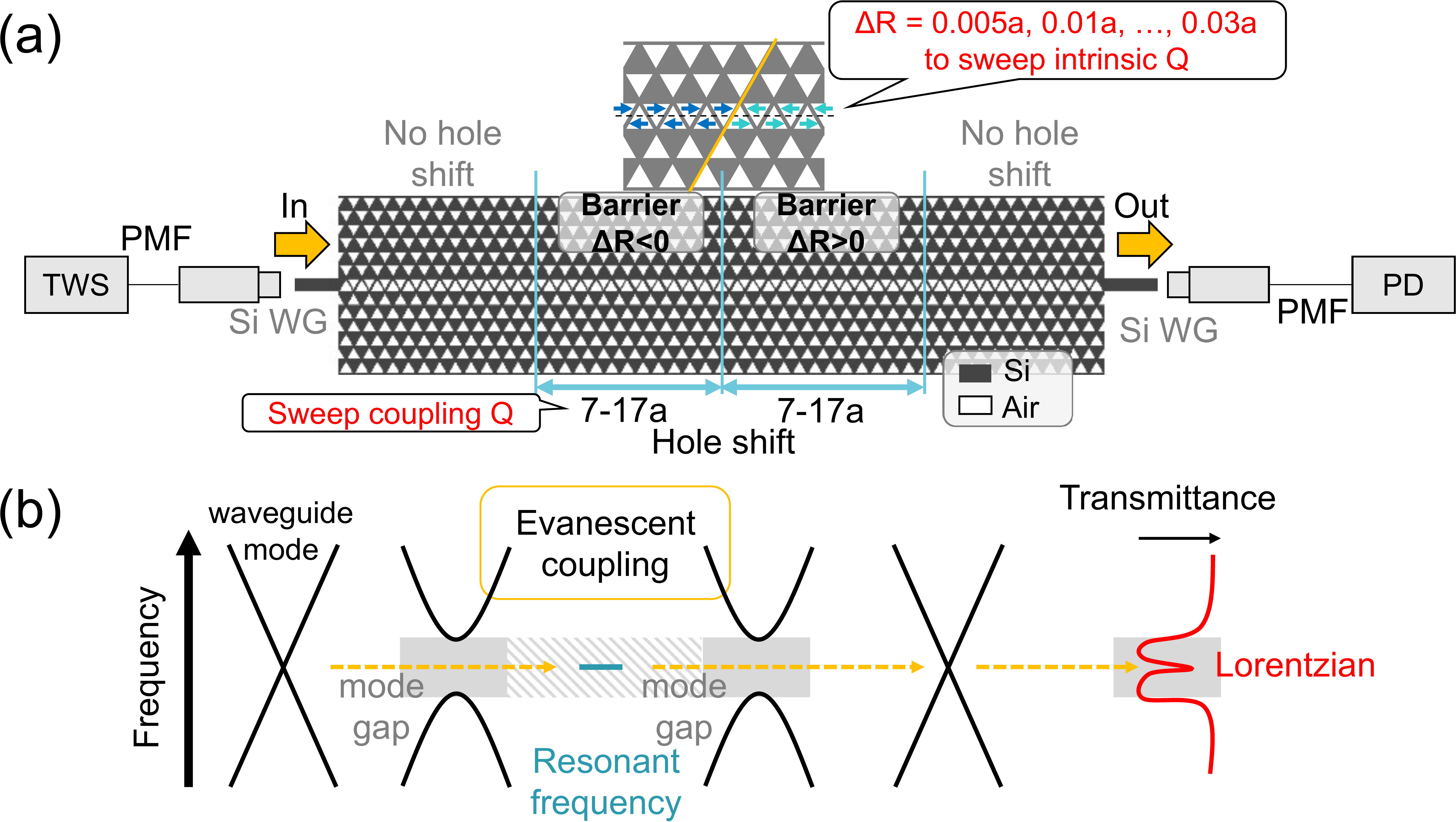}
	\caption{
			(a) Experimental setup for transmission measurements. A wavelength-tunable laser provides the input, and a photodiode detects the light collected from the output fiber. TWS, wavelength-tunable laser; PMF, polarization-maintaining fiber;
      PD, photodiode; and WG, waveguide.
			(b) Evanescent coupling between the propagating waveguide mode and the localized cavity mode through the shifted-hole barrier.
		}
	\label{fig:fig7}
\end{figure*}

The transmission setup is shown in Fig.~\ref{fig:fig7}(a). A wavelength-tunable laser covering $1360$--$1620\,\mathrm{nm}$ was used as a continuous-wave source, and $1\,\mathrm{mW}$ was delivered to the input fiber. A polarizer and lens at the tip of a polarization-maintaining single-mode fiber focused TE-polarized light onto the facet of the silicon access waveguide. The access waveguide was directly connected to the valley-photonic-crystal slab waveguide without a dedicated mode converter. The guided light propagated through the photonic-crystal waveguide and coupled evanescently to the nanocavity across the shifted-hole barrier.
The transmitted light was collected by an output fiber with the same optical arrangement and detected by a photodiode. The total length of the photonic-crystal section was $49.5a$. To vary the external coupling quality factor $Q_\mathrm{c}$, we fabricated six barrier lengths, corresponding to shifted-hole regions of $7a$, $9a$, $11a$, $13a$, $15a$, and $17a$ on each side of the nanocavity [Fig.~\ref{fig:fig7}(b)]. Six nonzero displacements, $\Delta R=0.005a$, $0.01a$, $0.015a$, $0.02a$, $0.025a$, and $0.03a$, were used to vary the intrinsic quality factor $Q_\mathrm{i}$ and mode volume. Thus, 36 perturbed devices covered all combinations of barrier length and displacement; unperturbed ($\Delta R=0$) reference structures were also measured.
Figure~\ref{fig:fig7}(b) illustrates the coupling mechanism. The shifted-hole regions act as barriers, so the propagating waveguide mode couples evanescently to the localized cavity mode. In this in-line geometry, resonant tunneling through the cavity produces a transmission peak near the cavity wavelength. The measured linewidth yields the loaded quality factor $Q_\mathrm{L}$, where $Q_\mathrm{L}^{-1}=Q_\mathrm{i}^{-1}+Q_\mathrm{c}^{-1}$.

\begin{figure}[htbp]
	\centering
	\includegraphics[width=\linewidth,height=0.74\textheight,keepaspectratio]{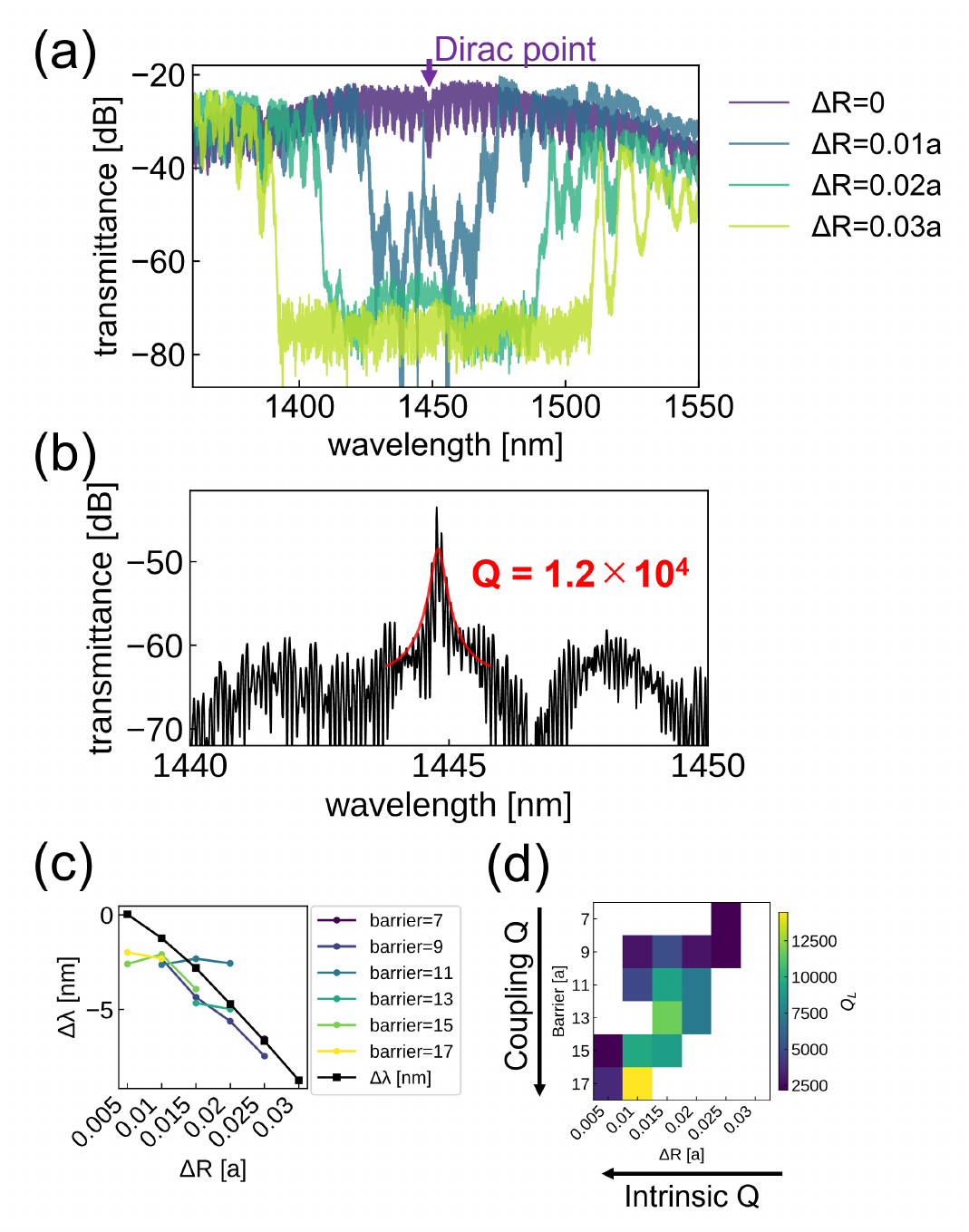}
	\caption{
  (a) Measured transmission spectra for several triangular-hole displacements $\Delta R$ at a fixed barrier length of $15a$.
  (b) Magnified spectrum and Lorentzian fit for $\Delta R=0.015a$ and a barrier length of $13a$.
  (c) Measured resonance shift as a function of $\Delta R$ for several barrier lengths, together with the simulated detuning $\Delta \lambda := \lambda_\mathrm{sim}-\lambda_\mathrm{dip}$. The simulated $\lambda_\mathrm{sim}-\lambda_\mathrm{Dirac}$ is also shown for comparison. 
  (d) Measured loaded-$Q$ map as a function of $\Delta R$ and barrier length. White cells indicate cases in which the resonance could not be identified reliably because of overlap with Fabry--P\'erot fringes or a low signal-to-noise ratio.
		}
	\label{fig:fig8}
\end{figure}

Figure~\ref{fig:fig8} summarizes the experimental characterization. Figure~\ref{fig:fig8}(a) shows transmission spectra for several displacements at a fixed barrier length of $15a$. For $\Delta R=0$, no cavity resonance is expected, and the spectrum is dominated by background transmission and Fabry--P\'erot fringes. A weak dip appears near $1450\,\mathrm{nm}$; below, we test its assignment as the Dirac-point reference by comparing the measured and simulated detunings. Introducing $\Delta R$ opens a mode gap around this wavelength. The measured gap widths are approximately $40$, $80$, and $120\,\mathrm{nm}$ for $\Delta R=0.01a$, $0.02a$, and $0.03a$, respectively, in reasonable agreement with Fig.~\ref{fig:fig5}(c). For $\Delta R=0.01a$, a narrow transmission peak appears within the gap, indicating resonant tunneling through the localized nanocavity mode.
Figure~\ref{fig:fig8}(b) shows the resonance for $\Delta R=0.015a$ and a barrier length of $13a$. A Lorentzian fit gives a loaded quality factor of $Q_\mathrm{L}=1.2\times10^4$. This narrow in-gap resonance is consistent with the localized mode created by the displacement-induced topological gap.

We next extracted the resonance shift for each barrier length [Fig.~\ref{fig:fig8}(c)]. In the simulation, the cavity detuning from the Dirac-point wavelength, $\lambda_\mathrm{sim}-\lambda_\mathrm{Dirac}$, becomes increasingly negative with $\Delta R$, corresponding to a systematic blueshift. The shift occurs because the present domain-wall state is not an exact zero mode of a perfectly chiral-symmetric SSH chain. The local domain-wall geometry introduces chiral-symmetry-breaking corrections, so the resonance is not pinned to the center of the gap. Increasing $\Delta R$ modifies both the interface-band dispersion and the local cavity field, thereby shifting the resonance. For each experimental series, we used the dip in the corresponding $\Delta R=0$ reference spectrum as the zero-detuning wavelength $\lambda_\mathrm{dip}$, and plot $\Delta \lambda := \lambda_\mathrm{sim}-\lambda_\mathrm{dip}$. Resonances that could not be identified reliably because of Fabry--P\'erot overlap or a low signal-to-noise ratio were excluded. The measured blueshift follows the simulated detuning, supporting---but not independently proving---the assignment of the reference dip to the glide-waveguide Dirac point. The dip may arise from a zero-index-like response near the Dirac point \cite{Huang2011}, disorder-induced scattering, or both.

Figure~\ref{fig:fig8}(d) shows the measured loaded quality factor $Q_\mathrm{L}$ as a function of $\Delta R$ and barrier length. White cells denote spectra for which the resonance could not be identified reliably. For short barriers, strong external coupling broadens the cavity resonance and allows it to overlap with Fabry--P\'erot fringes. For long barriers and large $\Delta R$, the stronger evanescent attenuation makes $Q_\mathrm{c}$ large and the transmitted resonance weak, leading to a poor signal-to-noise ratio. Within the observable region, $Q_\mathrm{L}$ tends to increase for longer barriers and smaller $\Delta R$, consistent with reduced external leakage and a larger intrinsic $Q_\mathrm{i}$. The maximum measured value is $Q_\mathrm{L}=1.2\times10^4$, below the ideal intrinsic value of approximately $10^5$ in Fig.~\ref{fig:fig5}. The reduction is consistent with fabrication disorder, surface roughness, and residual coupling loss commonly observed in photonic-crystal nanocavities \cite{Akahane2003,10.1063/1.2167801, Asano:17, Takata:23}. Further optimization should balance a high intrinsic $Q_\mathrm{i}$ against sufficient external coupling for reliable measurement. Additional spectra showing a clearly resolved resonance at the relatively large displacement $\Delta R=0.025a$ and the low-signal fitting uncertainty for $\Delta R=0.01a$ with a barrier length of $17a$ are presented in \ref{sec:nonoptimal_spectra}.
These results show that the observable coupling regime and loaded $Q$ factor can be engineered jointly through the barrier length and $\Delta R$. Together with Fig.~\ref{fig:fig5}(d,e), they demonstrate systematic control of both the spatial confinement and the optical linewidth of the nanocavity.

\begin{figure*}[htbp]
	\centering
	\includegraphics[width=0.9\linewidth]{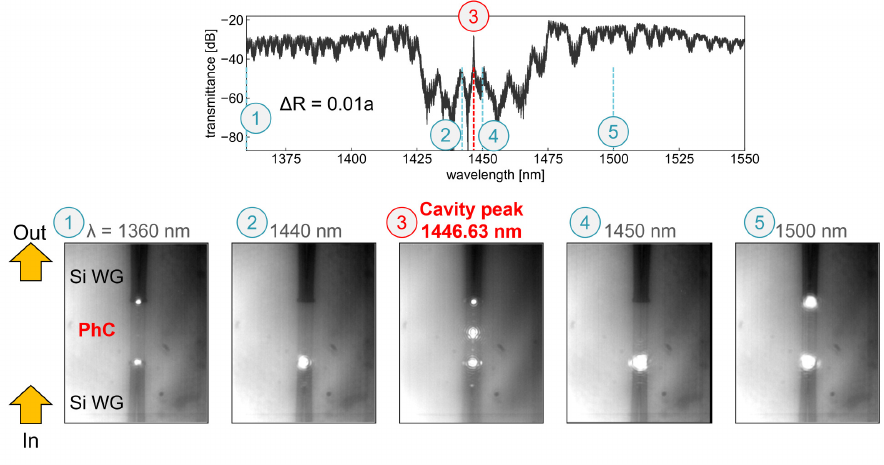}
	\caption{
			Top-view infrared images of light scattered from the device with a barrier length of $15a$ and $\Delta R=0.01a$, corresponding to Fig.~\ref{fig:fig8}(a). The five excitation wavelengths lie outside the mode gap (points 1 and 5), inside the gap but away from resonance (points 2 and 4), and at the nanocavity resonance (point 3, $\lambda=1446.63\,\mathrm{nm}$).
		}
	\label{fig:fig9}
\end{figure*}

To examine the spatial localization, we imaged light scattered from the top of the sample while injecting light through the lower silicon access waveguide [Fig.~\ref{fig:fig9}]. The measurements used the device with a barrier length of $15a$ and $\Delta R=0.01a$ at the five wavelengths marked in Fig.~\ref{fig:fig8}(a). Outside the mode gap (points 1 and 5), scattering is visible near both access-waveguide interfaces, consistent with propagation through the photonic-crystal waveguide. Inside the gap but away from resonance (points 2 and 4), scattering is concentrated near the input, consistent with suppressed propagation. At the resonance wavelength (point 3, $\lambda=1446.63\,\mathrm{nm}$), a pronounced scattering spot appears at the SSH domain wall in the center of the device. This spatially localized signal supports the formation of the domain-wall nanocavity mode.

\section{Conclusion}

We have theoretically and experimentally demonstrated a topological nanocavity formed by two orthogonal domain walls in a glide-symmetric valley photonic crystal. The valley-photonic-crystal interface provides transverse confinement, while an SSH-like domain wall introduced within the guided interface band provides longitudinal localization. Controlled displacement of the triangular holes opens a topological mode gap, and reversing the displacement produces waveguide sections with distinct Zak phases whose interface supports a localized cavity state. In fabricated silicon photonic-crystal slabs, we observed resonances within the induced mode gap and spatial localization at the domain wall, with a maximum measured loaded $Q$ factor of $1.2\times10^4$. We also demonstrated that the displacement amplitude $\Delta R$ provides continuous control over the mode gap and cavity confinement. 
This configuration provides a route to compact, continuously tunable cavity states integrated with valley-dependent optical responses, including polarization-selective excitation \cite{Sollner2015}, spin--momentum-locked light--matter interfaces, and slow-light propagation \cite{Yoshimi:20, Yoshimi:21}. Furthermore, because the cavity is introduced through a local perturbation of a glide-symmetric waveguide platform compatible with low-reflection bends \cite{Yoshimi:21, dai2023high}, the concept is well suited to embedding nanocavities in photonic circuits containing compact bends, junctions, and other routing elements.

\appendix
\def\thesection{Appendix \Alph{section}}

\section{Numerical method for calculating Zak phase}
\label{sec:zak_phase}

This appendix describes the numerical evaluation of the Zak phase \cite{PhysRevLett.62.2747} from the six-component electromagnetic Bloch eigenstates using a discretized Wilson-loop formulation based on Ref.~\cite{fukui2005chern}. For the lossless and nondispersive dielectric structure considered here, Maxwell's equations can be written as the generalized eigenproblem
\begin{align}
 c
 \begin{bmatrix}
  0 & -i\boldsymbol{\nabla}\times\\
  i\boldsymbol{\nabla}\times & 0
 \end{bmatrix}
 \widetilde{\Psi}_{n\boldsymbol{k}}(\boldsymbol{r})
 =\omega_{n\boldsymbol{k}}
 \begin{bmatrix}
  \varepsilon(\boldsymbol{r}) & 0\\
  0 & 1
 \end{bmatrix}
 \widetilde{\Psi}_{n\boldsymbol{k}}(\boldsymbol{r}),
 \label{eq:maxwell_six_component}
\end{align}
where $\varepsilon(\boldsymbol{r})$ is the relative permittivity and the full six-component Bloch state is
\begin{align}
 \widetilde{\Psi}_{n\boldsymbol{k}}(\boldsymbol{r})
 =\frac{1}{2\sqrt{U_{n\boldsymbol{k}}}}
 \begin{bmatrix}
  \sqrt{\varepsilon_0}\,\boldsymbol{\mathcal{E}}_{n\boldsymbol{k}}(\boldsymbol{r})\\
  \sqrt{\mu_0}\,\boldsymbol{\mathcal{H}}_{n\boldsymbol{k}}(\boldsymbol{r})
 \end{bmatrix}.
 \label{eq:full_six_component_state}
\end{align}
Here, $U_{n\boldsymbol{k}}$ is the time-averaged electromagnetic energy in the unit cell. The physical electric and magnetic Bloch fields are decomposed as
\begin{align}
 \boldsymbol{\mathcal{E}}_{n\boldsymbol{k}}(\boldsymbol{r})
 &=e^{i\boldsymbol{k}\cdot\boldsymbol{r}}
 \boldsymbol{E}_{n\boldsymbol{k}}(\boldsymbol{r}),\\
 \boldsymbol{\mathcal{H}}_{n\boldsymbol{k}}(\boldsymbol{r})
 &=e^{i\boldsymbol{k}\cdot\boldsymbol{r}}
 \boldsymbol{H}_{n\boldsymbol{k}}(\boldsymbol{r}),
 \label{eq:bloch_field_decomposition}
\end{align}
where $\boldsymbol{E}_{n\boldsymbol{k}}$ and $\boldsymbol{H}_{n\boldsymbol{k}}$ are periodic over the photonic-crystal unit cell. In the FEM calculation, a Bloch-periodic model was solved at each wavevector. The cell-periodic envelopes defined in Eq.~\eqref{eq:bloch_field_decomposition}, equivalently obtained by removing the factor $e^{i\boldsymbol{k}\cdot\boldsymbol{r}}$ from the periodic Bloch eigenfields, were used to evaluate the inter-wavevector overlaps. We therefore define the normalized cell-periodic six-component state as
\begin{align}
 \Psi_{n\boldsymbol{k}}(\boldsymbol{r})
 =\frac{1}{2\sqrt{U_{n\boldsymbol{k}}}}
 \begin{bmatrix}
  \sqrt{\varepsilon_0}\,\boldsymbol{E}_{n\boldsymbol{k}}(\boldsymbol{r})\\
  \sqrt{\mu_0}\,\boldsymbol{H}_{n\boldsymbol{k}}(\boldsymbol{r})
 \end{bmatrix},
 \label{eq:periodic_six_component_state}
\end{align}
The energy-weighted inner product is
\begin{align}
 \left\langle \boldsymbol{F}\middle|\boldsymbol{G}\right\rangle_B
 &:=\int_{\mathrm{u.c.}}\mathrm{d}^3\boldsymbol{r}\,
 \boldsymbol{F}^{\dagger}(\boldsymbol{r})
 B(\boldsymbol{r})
 \boldsymbol{G}(\boldsymbol{r}),
 \\
 B(\boldsymbol{r})
 &=
 \begin{bmatrix}
  \varepsilon(\boldsymbol{r}) & 0\\
  0 & 1
 \end{bmatrix},
 \label{eq:energy_weighted_inner_product}
\end{align}
so that $\langle\Psi_{n\boldsymbol{k}}|\Psi_{m\boldsymbol{k}}\rangle_B=\delta_{nm}$.

For an isolated band, the Zak phase is defined from the cell-periodic Bloch states by
\begin{align}
 \gamma_n
 =-i\int_{\mathrm{BZ}}
 \left\langle
  \Psi_{n\boldsymbol{k}}
  \middle|
  \frac{\partial}{\partial k_x}\Psi_{n\boldsymbol{k}}
 \right\rangle_B
 \mathrm{d}k_x
 \quad (\mathrm{mod}\ 2\pi).
 \label{eq:zak_phase_continuous}
\end{align}
We discretize the Brillouin zone as
\begin{align}
 \boldsymbol{k}_j
 &=\boldsymbol{k}_{\mathrm{s}}+j\Delta\boldsymbol{k},
 \qquad j=0,1,\ldots,N,
 \\
 \boldsymbol{k}_N
 &=\boldsymbol{k}_{\mathrm{s}}+\boldsymbol{G},
 \label{eq:zak_k_path}
\end{align}
where $\boldsymbol{G}$ is the reciprocal-lattice vector along the waveguide. For neighboring wavevectors, the overlap matrix element is
\begin{align}
 M_n^{(j)}
 &:={\left\langle
  \Psi_{n\boldsymbol{k}_j}
  \middle|
  \Psi_{n\boldsymbol{k}_{j+1}}
 \right\rangle}_B
 \nonumber\\
 &=\int_{\mathrm{u.c.}}\mathrm{d}^3\boldsymbol{r}\,
 \frac{1}{4\sqrt{U_{n\boldsymbol{k}_j}U_{n\boldsymbol{k}_{j+1}}}}
 \Bigl[
  \varepsilon_0\varepsilon(\boldsymbol{r})
  \boldsymbol{E}^{*}_{n\boldsymbol{k}_j}(\boldsymbol{r})\cdot
  \boldsymbol{E}_{n\boldsymbol{k}_{j+1}}(\boldsymbol{r})\nonumber\\
  &\qquad \qquad \qquad \qquad \qquad +\mu_0
  \boldsymbol{H}^{*}_{n\boldsymbol{k}_j}(\boldsymbol{r})\cdot
  \boldsymbol{H}_{n\boldsymbol{k}_{j+1}}(\boldsymbol{r})
 \Bigr].
 \label{eq:zak_overlap}
\end{align}
We retain only the phase of each overlap by defining the link variable
\begin{align}
 \mathcal{U}_n^{(j)}
 :=\frac{M_n^{(j)}}{|M_n^{(j)}|}.
 \label{eq:zak_link_variable}
\end{align}
The loop is closed using the reciprocal-lattice sewing relation
\begin{align}
 \Psi_{n,\boldsymbol{k}_{\mathrm{s}}+\boldsymbol{G}}(\boldsymbol{r})
 =e^{-i\boldsymbol{G}\cdot\boldsymbol{r}}
 \Psi_{n\boldsymbol{k}_{\mathrm{s}}}(\boldsymbol{r}),
 \label{eq:zak_sewing_relation}
\end{align}
up to an arbitrary overall phase. The discretized Zak phase is then
\begin{align}
 \gamma_n
 =\operatorname{Im}
 \left[
  \log\prod_{j=0}^{N-1}\mathcal{U}_n^{(j)}
 \right]
 \quad (\mathrm{mod}\ 2\pi)
 \label{eq:zak_phase_numerical}
\end{align}
which is invariant under independent phase choices for the numerical eigenmode at each wavevector. In the calculations shown in Fig.~\ref{fig:fig2}(b), we used $\boldsymbol{k}_{\mathrm{s}}=0$, $\boldsymbol{G}=[2\pi/a,0,0]^\top$, and $N=40$. The individual-band Zak phase was evaluated only for $\Delta R\neq0$, because the upper and lower interface bands become degenerate at $\Delta R=0$ and are then not isolated over the entire Brillouin zone.

\section{Derivation of the effective Hamiltonian near the Dirac point}
\label{sec:kp_method}

This section derives the two-band effective Hamiltonian by projecting the Hermitian Maxwell operator onto the degenerate guided modes at the Dirac point.  The derivation is written for a three-dimensional vector field and therefore also applies to a finite-thickness photonic-crystal slab.

We write the magnetic Bloch mode as
\begin{align}
 \boldsymbol{H}_{n\boldsymbol{k}}(\boldsymbol{r})
 =e^{i\boldsymbol{k}\cdot\boldsymbol{r}}
 \boldsymbol{\varphi}_{n\boldsymbol{k}}(\boldsymbol{r}),
\end{align}
where \(\boldsymbol{\varphi}_{n\boldsymbol{k}}\) is lattice periodic.  For a real, positive relative permittivity, define
\begin{align}
 \eta(\boldsymbol{r})
 :=\frac{1}{\varepsilon_{\mathrm{p}}(\boldsymbol{r})},
 \qquad
 \hat{C}_{\boldsymbol{k}}
 :=\left(\boldsymbol{\nabla}+i\boldsymbol{k}\right)\times ,
\end{align}
where \(\varepsilon_{\mathrm{p}}\) denotes the permittivity of the perturbed structure.  The Maxwell eigenproblem for the periodic magnetic field is
\begin{align}
 \hat{\Theta}(\boldsymbol{k},\eta)
 \boldsymbol{\varphi}_{n\boldsymbol{k}}
 :=\hat{C}_{\boldsymbol{k}}\eta\hat{C}_{\boldsymbol{k}}
 \boldsymbol{\varphi}_{n\boldsymbol{k}}
 =\lambda_{n\boldsymbol{k}}\boldsymbol{\varphi}_{n\boldsymbol{k}},
 \qquad
 \lambda_{n\boldsymbol{k}}=\left(\frac{\omega_{n\boldsymbol{k}}}{c}\right)^2 .
 \label{eq:theta_bloch}
\end{align}
With the standard inner product
\(\langle\boldsymbol{f}|\boldsymbol{g}\rangle
=\int_{\mathrm{u.c.}}\boldsymbol{f}^{*}\cdot\boldsymbol{g}\,\mathrm{d}^{3}r\),
\(\hat{C}_{\boldsymbol{k}}\) is Hermitian under Bloch-periodic boundary conditions.  Consequently, \(\hat{\Theta}(\boldsymbol{k},\eta)\) is Hermitian for real \(\eta\).  Any finite-dimensional Galerkin projection of this operator is therefore Hermitian as well.

Let \(\boldsymbol{k}_{0}=k_{0}\hat{\boldsymbol{x}}\), with \(k_{0}=\pi/a\), and let \(\boldsymbol{\varphi}^{(0)}_{\mathrm{u}}\) and \(\boldsymbol{\varphi}^{(0)}_{\mathrm{d}}\) be the two orthonormal eigenmodes at the unperturbed Dirac point:
\begin{align}
 \hat{\Theta}_{0}\boldsymbol{\varphi}^{(0)}_{\alpha}
 =\lambda_{D}\boldsymbol{\varphi}^{(0)}_{\alpha},
 \qquad
 \left\langle\boldsymbol{\varphi}^{(0)}_{\alpha}
 \middle|
 \boldsymbol{\varphi}^{(0)}_{\beta}\right\rangle
 =\delta_{\alpha\beta},
 \qquad
 \alpha,\beta\in\{\mathrm{u},\mathrm{d}\},
 \label{eq:dirac_basis}
\end{align}
where \(\lambda_{D}=(\omega_{D}/c)^2\) and
\(\hat{\Theta}_{0}=\hat{C}_{0}\eta_{0}\hat{C}_{0}\),
\(\eta_{0}=1/\varepsilon\), and
\(\hat{C}_{0}=(\boldsymbol{\nabla}+ik_{0}\hat{\boldsymbol{x}})\times\). Then, 
we introduce the wavevector displacement \(q:=\delta k_{x}=k_{x}-k_{0}\) and the inverse-permittivity perturbation
\begin{align}
 \Delta\eta(\boldsymbol{r})
 :=\frac{1}{\varepsilon_{\mathrm{p}}(\boldsymbol{r})}
   -\frac{1}{\varepsilon(\boldsymbol{r})}.
 \label{eq:delta_eta}
\end{align}
Writing
\(\hat{C}_{\boldsymbol{k}}=\hat{C}_{0}+q\hat{C}_{x}\), with
\(\hat{C}_{x}:=i\hat{\boldsymbol{x}}\times\), gives
\begin{align}
 \hat{\Theta}(\boldsymbol{k},\eta)
 &=\hat{\Theta}_{0}
   +q\hat{V}_{k}
   +\hat{V}_{\varepsilon}
   +\mathcal{O}\!\left(q^{2},\Delta\eta^{2},q\Delta\eta\right),
 \label{eq:theta_expansion}\\
 \hat{V}_{k}
 &:=\hat{C}_{x}\eta_{0}\hat{C}_{0}
   +\hat{C}_{0}\eta_{0}\hat{C}_{x},
 \qquad
 \hat{V}_{\varepsilon}
 :=\hat{C}_{0}\Delta\eta\hat{C}_{0}.
 \label{eq:vk_veps}
\end{align}
Equation~\eqref{eq:theta_expansion} retains terms that are first order in the joint expansion about \((q,\Delta\eta)=(0,0)\).  The mixed correction proportional to \(q\Delta\eta\) is discussed below.
Expanding a mode near the degeneracy as
\(\boldsymbol{\varphi}_{n\boldsymbol{k}}
=\sum_{\beta}C_{n\beta}\boldsymbol{\varphi}^{(0)}_{\beta}\)
and projecting Eq.~\eqref{eq:theta_bloch} onto the two Dirac-point states yields
\begin{align}
 \sum_{\beta}
 \left[
  \lambda_{D}\delta_{\alpha\beta}
  +qP_{\alpha\beta}
  +M_{\alpha\beta}
 \right]C_{n\beta}
 =\lambda_{n\boldsymbol{k}}C_{n\alpha},
 \label{eq:projected_two_band}
\end{align}
where
\begin{align}
 P_{\alpha\beta}
 &:=\left\langle\boldsymbol{\varphi}^{(0)}_{\alpha}
 \middle|\hat{V}_{k}\middle|
 \boldsymbol{\varphi}^{(0)}_{\beta}\right\rangle,
 \label{eq:p_operator}\\
 M_{\alpha\beta}
 &:=\left\langle\boldsymbol{\varphi}^{(0)}_{\alpha}
 \middle|\hat{V}_{\varepsilon}\middle|
 \boldsymbol{\varphi}^{(0)}_{\beta}\right\rangle.
 \label{eq:m_operator}
\end{align}
Both matrices are Hermitian:
\(P_{\beta\alpha}=P_{\alpha\beta}^{*}\) and
\(M_{\beta\alpha}=M_{\alpha\beta}^{*}\).

To express these matrix elements in terms of the electric and magnetic fields, we use the unperturbed Maxwell relation
\begin{align}
 \hat{C}_{0}\boldsymbol{\varphi}^{(0)}_{\alpha}
 =-i\varepsilon(\boldsymbol{r})\frac{\omega_{D}}{c}
 \boldsymbol{\psi}^{(0)}_{\alpha}(\boldsymbol{r}),
 \label{eq:curl_h_to_e}
\end{align}
where \(\boldsymbol{\psi}^{(0)}_{\alpha}\) is the periodic electric-field basis function.  Substitution into Eqs.~\eqref{eq:p_operator} and \eqref{eq:m_operator} gives
\begin{align}
 P_{\alpha\beta}
 &=\frac{\omega_{D}}{c}
 \int_{\mathrm{u.c.}}
 \left[
  \boldsymbol{\psi}^{(0)*}_{\alpha}\times
  \boldsymbol{\varphi}^{(0)}_{\beta}
  -
  \boldsymbol{\varphi}^{(0)*}_{\alpha}\times
  \boldsymbol{\psi}^{(0)}_{\beta}
 \right]_{x}
 \mathrm{d}^{3}r,
 \label{eq:p_field}\\
 M_{\alpha\beta}
 &=\frac{\omega_{D}^{2}}{c^{2}}
 \int_{\mathrm{u.c.}}
 \varepsilon^{2}(\boldsymbol{r})\Delta\eta(\boldsymbol{r})
 \boldsymbol{\psi}^{(0)*}_{\alpha}(\boldsymbol{r})\cdot
 \boldsymbol{\psi}^{(0)}_{\beta}(\boldsymbol{r})
 \mathrm{d}^{3}r.
 \label{eq:m_field_exact}
\end{align}
If the permittivity contrast itself is treated as a small perturbation,
\(\varepsilon_{\mathrm{p}}=\varepsilon+\Delta\varepsilon\), then
\begin{align}
 \varepsilon^{2}\Delta\eta
 =-\Delta\varepsilon+\mathcal{O}(\Delta\varepsilon^{2}),
\end{align}
and Eq.~\eqref{eq:m_field_exact} reduces to
\begin{align}
 M_{\alpha\beta}
 \simeq
 -\frac{\omega_{D}^{2}}{c^{2}}
 \int_{\mathrm{u.c.}}
 \Delta\varepsilon(\boldsymbol{r})
 \boldsymbol{\psi}^{(0)*}_{\alpha}(\boldsymbol{r})\cdot
 \boldsymbol{\psi}^{(0)}_{\beta}(\boldsymbol{r})
 \mathrm{d}^{3}r.
 \label{eq:m_field_weak}
\end{align}
For the symmetry-adapted standing-wave basis used in Fig.~\ref{fig:fig3}, the mirror/glide and time-reversal constraints permit a phase convention in which
\begin{align}
 P_{\mathrm{u}\mathrm{u}}=P_{\mathrm{d}\mathrm{d}}=0,
 \quad
 P_{\mathrm{u}\mathrm{d}}=iv_{\mathrm{D}},
 \quad
 P_{\mathrm{d}\mathrm{u}}=-iv_{\mathrm{D}},
 \quad
 M_{\mathrm{u}\mathrm{d}}=M_{\mathrm{d}\mathrm{u}}\in\mathbb{R},
 \label{eq:p_symmetry_form}
\end{align}
where $v_{\mathrm{D}}=-iP_{\mathrm{u}\mathrm{d}}\in\mathbb{R}$. The remaining constant Hermitian matrix is decomposed as
\begin{align}
 \overline{M}
 &:=\frac{M_{\mathrm{u}\mathrm{u}}+M_{\mathrm{d}\mathrm{d}}}{2},
 &
 \kappa
 &:=M_{\mathrm{u}\mathrm{d}}\in\mathbb{R},
 &
 m_{\mathrm{D}}
 &:=\frac{M_{\mathrm{u}\mathrm{u}}-M_{\mathrm{d}\mathrm{d}}}{2}.
 \label{eq:mass_definitions}
\end{align}
After subtracting the scalar shift \((\lambda_{D}+\overline{M})I_{2}\), Eq.~\eqref{eq:projected_two_band} becomes
\begin{align}
 H'
 &=
 \begin{bmatrix}
  m_{\mathrm{D}} & \kappa+iv_{\mathrm{D}}q\\
  \kappa-iv_{\mathrm{D}}q & -m_{\mathrm{D}}
 \end{bmatrix}
 =\kappa\sigma_{x}-v_{\mathrm{D}}q\sigma_{y}
  +m_{\mathrm{D}}\sigma_{z}.
 \label{eq:effective_hamiltonian_appendix}
\end{align}
This is the Hermitian two-band Hamiltonian quoted in the main text.  In the weak-permittivity-perturbation form, the two gap parameters are
\begin{align}
 \kappa
 &\simeq
 -\frac{\omega_{D}^{2}}{c^{2}}
 \int_{\mathrm{u.c.}}
 \Delta\varepsilon
 \boldsymbol{\psi}^{(0)*}_{\mathrm{u}}\cdot
 \boldsymbol{\psi}^{(0)}_{\mathrm{d}}\,\mathrm{d}^{3}r,
 \label{eq:kappa_corrected}\\
 m_{\mathrm{D}}
 &\simeq
 \frac{\omega_{D}^{2}}{2c^{2}}
 \int_{\mathrm{u.c.}}
 \Delta\varepsilon
 \left(
  \left|\boldsymbol{\psi}^{(0)}_{\mathrm{d}}\right|^{2}
  -
  \left|\boldsymbol{\psi}^{(0)}_{\mathrm{u}}\right|^{2}
 \right)\mathrm{d}^{3}r.
 \label{eq:md_corrected}
\end{align}
%
The sign of \(\kappa\) reverses when the displacement is reversed; its association with the Zak phase depends on the unit-cell and basis conventions and is fixed by the numerical Zak-phase calculation.

The combined glide--mirror symmetry exchanges the intensities of the \(\mathrm{u}\) and \(\mathrm{d}\) states.  If the scalar weight
\(\varepsilon^{2}\Delta\eta\) in Eq.~\eqref{eq:m_field_exact} is invariant under this operation, then
\(M_{\mathrm{u}\mathrm{u}}=M_{\mathrm{d}\mathrm{d}}\) and hence \(m_{\mathrm{D}}=0\).  The traceless Hamiltonian then contains only \(\sigma_{x}\) and \(\sigma_{y}\) terms and anticommutes with \(\sigma_{z}\), reproducing the local chiral form of the SSH Hamiltonian.

\section{Additional representative transmission spectra}
\label{sec:nonoptimal_spectra}

\begin{figure}[htbp]
	\centering
	\includegraphics[width=1\linewidth]{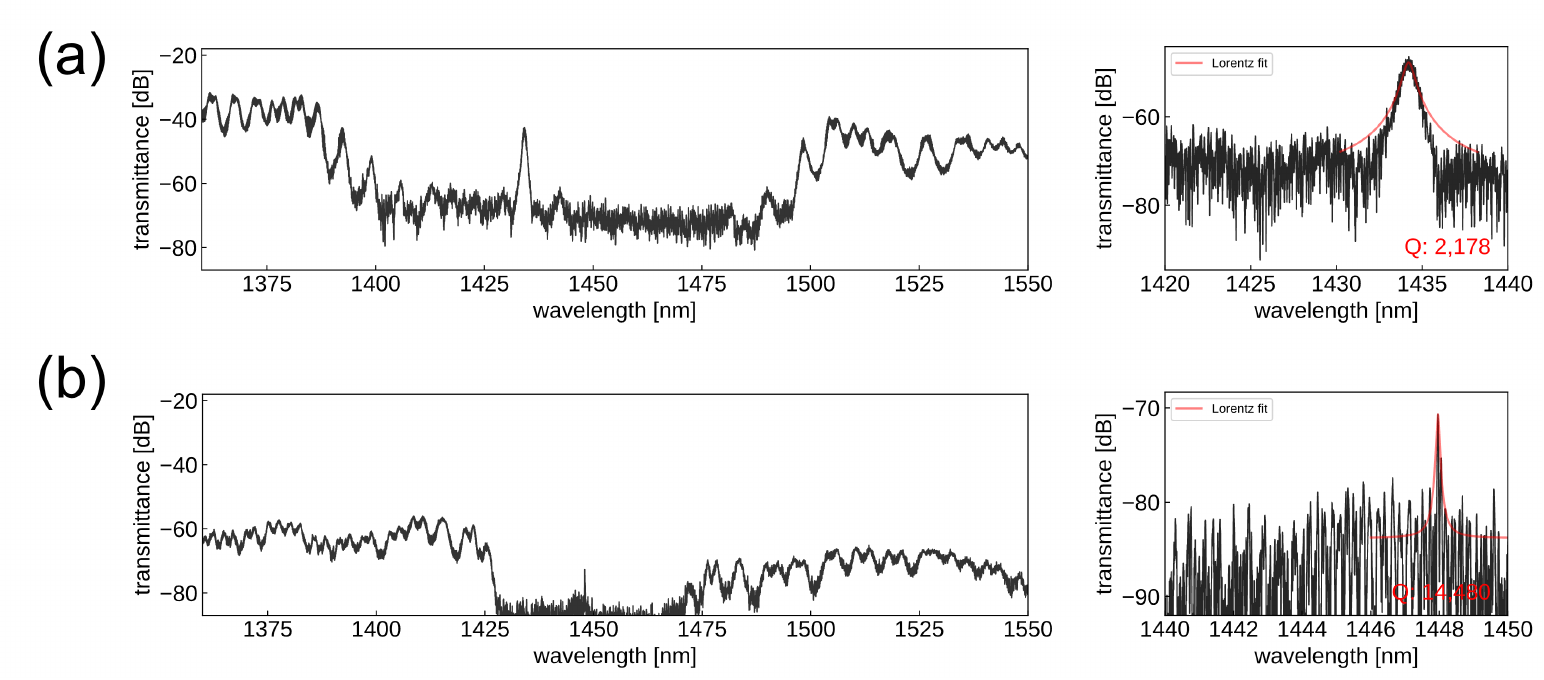}
	\caption{
		Representative measured transmission spectra for non-optimal coupling conditions: (a) $\Delta R = 0.025a$ with a barrier length of $7a$, and (b) $\Delta R = 0.01a$ with a barrier length of $17a$.
		}
	\label{fig:figS1}
\end{figure}

Figure~\ref{fig:figS1} shows two additional transmission spectra that complement the results presented in the main text. The broadband transmission spectra were acquired with a wavelength sampling interval of $10\,\mathrm{pm}$, while separate high-resolution spectra around the cavity resonances were acquired with a sampling interval of $1\,\mathrm{pm}$.

Figure~\ref{fig:figS1}(a), measured at $\Delta R=0.025a$ with a barrier length of $7a$, shows a well-resolved transmission peak within the mode gap. A Lorentzian fit gives a loaded quality factor of $Q_\mathrm{L}=2.2\times10^3$. This value is consistent with the calculated intrinsic quality factor of $Q_\mathrm{i}=4.4\times10^3$ at this displacement, because the measured $Q_\mathrm{L}$ also includes the loss due to external coupling. The clear peak demonstrates that the cavity resonance remains observable at this relatively large displacement. Together with the numerical result that increasing $\Delta R$ reduces the mode volume while lowering the radiative $Q$ factor [Fig.~\ref{fig:fig5}(d)], this observation provides additional support for tuning the balance between the $Q$ factor and mode volume through $\Delta R$.

Figure~\ref{fig:figS1}(b) shows a spectrum measured at $\Delta R=0.01a$ with a barrier length of $17a$. A defect in the free-space-to-chip coupling region reduced the transmission outside the mode gap to approximately $-60\,\mathrm{dB}$. The gap and resonance remain visible, and a Lorentzian fit gives a nominal loaded quality factor of $Q_\mathrm{L}=1.4\times10^4$; however, the low signal level increases the fitting uncertainty, and this fitted value should therefore be interpreted with caution.

\section*{Funding}
This work was supported by the Japan Society for the Promotion of Science (Grant Nos. JP20H05641, JP21K14551, 24K01377, 24H02232, and 24H00400) and JST Support for Pioneering Research Initiated by the Next Generation (JPMJSP2180).

\section*{Acknowledgements}
We thank Dr. Toshiaki Tamamura, Toshifumi Watanabe, Osamu Moriwaki, and Junichi Asaoka for their contributions to the fabrication process. We also thank Dr. Kenta Takata for fruitful discussions.

\section*{Disclosures}
The authors declare no conflicts of interest.

\section*{Data availability}
Data underlying the results presented in this paper are not publicly available at this time but may be obtained from the authors upon reasonable request.

\bibliographystyle{apsrev4-2}
\bibliography{wgssh}

@article{PhysRevLett.42.1698,
  title = {Solitons in Polyacetylene},
  author = {Su, W. P. and Schrieffer, J. R. and Heeger, A. J.},
  journal = {Phys. Rev. Lett.},
  volume = {42},
  issue = {25},
pages = {1698--1701},
  numpages = {0},
  year = {1979},
  month = {Jun},
  publisher = {American Physical Society},
  doi = {10.1103/PhysRevLett.42.1698},
  url = {https://link.aps.org/doi/10.1103/PhysRevLett.42.1698}
}

@book{asboth2016short,
  author    = {Asb{\'o}th, J{\'a}nos K. and Oroszl{\'a}ny, L{\'a}szl{\'o} and P{\'a}lyi, Andr{\'a}s},
  title     = {A Short Course on Topological Insulators: Band Structure and Edge States in One and Two Dimensions},
  series    = {Lecture Notes in Physics},
  volume    = {919},
  publisher = {Springer},
  address   = {Cham},
  year      = {2016},
  doi       = {10.1007/978-3-319-25607-8},
  url       = {https://doi.org/10.1007/978-3-319-25607-8}
}

@Article{Meier2016,
author={Meier, Eric J.
and An, Fangzhao Alex
and Gadway, Bryce},
title={Observation of the topological soliton state in the Su--Schrieffer--Heeger model},
journal={Nature Communications},
year={2016},
month={Dec},
day={23},
volume={7},
number={1},
pages = {13986},
issn={2041-1723},
doi={10.1038/ncomms13986},
url={https://doi.org/10.1038/ncomms13986}
}

@Article{Lohse2016,
author={Lohse, M.
and Schweizer, C.
and Zilberberg, O.
and Aidelsburger, M.
and Bloch, I.},
title={A Thouless quantum pump with ultracold bosonic atoms in an optical superlattice},
journal={Nature Physics},
year={2016},
month={Apr},
day={01},
volume={12},
number={4},
pages = {350--354},
issn={1745-2481},
doi={10.1038/nphys3584},
url={https://doi.org/10.1038/nphys3584}
}

@Article{Atala2013,
author={Atala, Marcos
and Aidelsburger, Monika
and Barreiro, Julio T.
and Abanin, Dmitry
and Kitagawa, Takuya
and Demler, Eugene
and Bloch, Immanuel},
title={Direct measurement of the Zak phase in topological Bloch bands},
journal={Nature Physics},
year={2013},
month={Dec},
day={01},
volume={9},
number={12},
pages = {795--800},
issn={1745-2481},
doi={10.1038/nphys2790},
url={https://doi.org/10.1038/nphys2790}
}

@article{MoritakeOnoNotomi+2022+2183+2189,
url = {https://doi.org/10.1515/nanoph-2021-0648},
title = {Far-field optical imaging of topological edge states in zigzag plasmonic chains},
author = {Yuto Moritake and Masaaki Ono and Masaya Notomi},
pages = {2183--2189},
volume = {11},
number = {9},
journal = {Nanophotonics},
doi = {10.1515/nanoph-2021-0648},
year = {2022},
lastchecked = {2025-01-16}
}

@article{doi:10.1126/sciadv.abf8049,
author = {Rui Su  and Sanjib Ghosh  and Timothy C. H. Liew  and Qihua Xiong },
title = {Optical switching of topological phase in a perovskite polariton lattice},
journal = {Science Advances},
volume = {7},
number = {21},
pages = {eabf8049},
year = {2021},
doi = {10.1126/sciadv.abf8049},
URL = {https://www.science.org/doi/abs/10.1126/sciadv.abf8049},
eprint = {https://www.science.org/doi/pdf/10.1126/sciadv.abf8049}}

@article{PhysRevLett.120.113901,
  title = {Edge-Mode Lasing in {1D} Topological Active Arrays},
  author = {Parto, Midya and Wittek, Steffen and Hodaei, Hossein and Harari, Gal and Bandres, Miguel A. and Ren, Jinhan and Rechtsman, Mikael C. and Segev, Mordechai and Christodoulides, Demetrios N. and Khajavikhan, Mercedeh},
  journal = {Phys. Rev. Lett.},
  volume = {120},
  issue = {11},
pages = {113901},
  numpages = {6},
  year = {2018},
  month = {Mar},
  publisher = {American Physical Society},
  doi = {10.1103/PhysRevLett.120.113901},
  url = {https://link.aps.org/doi/10.1103/PhysRevLett.120.113901}
}

@Article{St-Jean2017,
author={St-Jean, P.
and Goblot, V.
and Galopin, E.
and Lema{\^i}tre, A.
and Ozawa, T.
and Le Gratiet, L.
and Sagnes, I.
and Bloch, J.
and Amo, A.},
title={Lasing in topological edge states of a one-dimensional lattice},
journal={Nature Photonics},
year={2017},
month={Oct},
day={01},
volume={11},
number={10},
pages = {651--656},
issn={1749-4893},
doi={10.1038/s41566-017-0006-2},
url={https://doi.org/10.1038/s41566-017-0006-2}
}

@Article{Zhao2018,
author={Zhao, Han
and Miao, Pei
and Teimourpour, Mohammad H.
and Malzard, Simon
and El-Ganainy, Ramy
and Schomerus, Henning
and Feng, Liang},
title={Topological hybrid silicon microlasers},
journal={Nature Communications},
year={2018},
month={Mar},
day={07},
volume={9},
number={1},
pages = {981},
issn={2041-1723},
doi={10.1038/s41467-018-03434-2},
url={https://doi.org/10.1038/s41467-018-03434-2}
}

@Article{Han2019,
author={Han, Changhyun
and Lee, Myungjae
and Callard, S{\'e}gol{\`e}ne
and Seassal, Christian
and Jeon, Heonsu},
title={Lasing at topological edge states in a photonic crystal {L3} nanocavity dimer array},
journal={Light: Science {\&} Applications},
year={2019},
month={Apr},
day={24},
volume={8},
number={1},
pages = {40},
issn={2047-7538},
doi={10.1038/s41377-019-0149-7},
url={https://doi.org/10.1038/s41377-019-0149-7}
}

@article{PhysRevLett.114.223901,
  title = {Scheme for Achieving a Topological Photonic Crystal by Using Dielectric Material},
  author = {Wu, Long-Hua and Hu, Xiao},
  journal = {Phys. Rev. Lett.},
  volume = {114},
  issue = {22},
pages = {223901},
  numpages = {5},
  year = {2015},
  month = {Jun},
  publisher = {American Physical Society},
  doi = {10.1103/PhysRevLett.114.223901},
  url = {https://link.aps.org/doi/10.1103/PhysRevLett.114.223901}
}

@article{doi:10.1126/science.aaq0327,
author = {Sabyasachi Barik  and Aziz Karasahin  and Christopher Flower  and Tao Cai  and Hirokazu Miyake  and Wade DeGottardi  and Mohammad Hafezi  and Edo Waks },
title = {A topological quantum optics interface},
journal = {Science},
volume = {359},
number = {6376},
pages = {666--668},
year = {2018},
doi = {10.1126/science.aaq0327},
URL = {https://www.science.org/doi/abs/10.1126/science.aaq0327},
eprint = {https://www.science.org/doi/pdf/10.1126/science.aaq0327}}

@article{doi:10.1126/sciadv.aaw4137,
author = {Nikhil Parappurath  and Filippo Alpeggiani  and L. Kuipers  and Ewold Verhagen },
title = {Direct observation of topological edge states in silicon photonic crystals: Spin, dispersion, and chiral routing},
journal = {Science Advances},
volume = {6},
number = {10},
pages = {eaaw4137},
year = {2020},
doi = {10.1126/sciadv.aaw4137},
URL = {https://www.science.org/doi/abs/10.1126/sciadv.aaw4137},
eprint = {https://www.science.org/doi/pdf/10.1126/sciadv.aaw4137}}

@Article{Dong2017,
author={Dong, Jian-Wen
and Chen, Xiao-Dong
and Zhu, Hanyu
and Wang, Yuan
and Zhang, Xiang},
title={Valley photonic crystals for control of spin and topology},
journal={Nature Materials},
year={2017},
month={Mar},
day={01},
volume={16},
number={3},
pages = {298--302},
issn={1476-4660},
doi={10.1038/nmat4807},
url={https://doi.org/10.1038/nmat4807}
}

@Article{Shalaev2019,
author={Shalaev, Mikhail I.
and Walasik, Wiktor
and Tsukernik, Alexander
and Xu, Yun
and Litchinitser, Natalia M.},
title={Robust topologically protected transport in photonic crystals at telecommunication wavelengths},
journal={Nature Nanotechnology},
year={2019},
month={Jan},
day={01},
volume={14},
number={1},
pages = {31--34},
issn={1748-3395},
doi={10.1038/s41565-018-0297-6},
url={https://doi.org/10.1038/s41565-018-0297-6}
}

@Article{Arora2021,
author={Arora, Sonakshi
and Bauer, Thomas
and Barczyk, Ren{\'e}
and Verhagen, Ewold
and Kuipers, L.},
title={Direct quantification of topological protection in symmetry-protected photonic edge states at telecom wavelengths},
journal={Light: Science {\&} Applications},
year={2021},
month={Jan},
day={06},
volume={10},
number={1},
pages = {9},
issn={2047-7538},
doi={10.1038/s41377-020-00458-6},
url={https://doi.org/10.1038/s41377-020-00458-6}
}

@article{Yoshimi:20,
author = {Hironobu Yoshimi and Takuto Yamaguchi and Yasutomo Ota and Yasuhiko Arakawa and Satoshi Iwamoto},
journal = {Opt. Lett.},
number = {9},
pages = {2648--2651},
publisher = {Optica Publishing Group},
title = {Slow light waveguides in topological valley photonic crystals},
volume = {45},
month = {May},
year = {2020},
url = {https://opg.optica.org/ol/abstract.cfm?URI=ol-45-9-2648},
doi = {10.1364/OL.391764},
}

@article{Yoshimi:21,
author = {Hironobu Yoshimi and Takuto Yamaguchi and Ryota Katsumi and Yasutomo Ota and Yasuhiko Arakawa and Satoshi Iwamoto},
journal = {Opt. Express},
number = {9},
pages = {13441--13450},
publisher = {Optica Publishing Group},
title = {Experimental demonstration of topological slow light waveguides in valley photonic crystals},
volume = {29},
month = {Apr},
year = {2021},
url = {https://opg.optica.org/oe/abstract.cfm?URI=oe-29-9-13441},
doi = {10.1364/OE.422962},
}

@article{Ota:19,
author = {Yasutomo Ota and Feng Liu and Ryota Katsumi and Katsuyuki Watanabe and Katsunori Wakabayashi and Yasuhiko Arakawa and Satoshi Iwamoto},
journal = {Optica},
number = {6},
pages = {786--789},
publisher = {Optica Publishing Group},
title = {Photonic crystal nanocavity based on a topological corner state},
volume = {6},
month = {Jun},
year = {2019},
url = {https://opg.optica.org/optica/abstract.cfm?URI=optica-6-6-786},
doi = {10.1364/OPTICA.6.000786},
}

@Article{Ota2018,
author={Ota, Yasutomo
and Katsumi, Ryota
and Watanabe, Katsuyuki
and Iwamoto, Satoshi
and Arakawa, Yasuhiko},
title={Topological photonic crystal nanocavity laser},
journal={Communications Physics},
year={2018},
month={Nov},
day={22},
volume={1},
number={1},
pages = {86},
issn={2399-3650},
doi={10.1038/s42005-018-0083-7},
url={https://doi.org/10.1038/s42005-018-0083-7}
}

@Article{Akahane2003,
author={Akahane, Yoshihiro
and Asano, Takashi
and Song, Bong-Shik
and Noda, Susumu},
title={High-Q photonic nanocavity in a two-dimensional photonic crystal},
journal={Nature},
year={2003},
month={Oct},
day={01},
volume={425},
number={6961},
pages = {944--947},
issn={1476-4687},
doi={10.1038/nature02063},
url={https://doi.org/10.1038/nature02063}
}

@article{10.1063/1.2167801,
    author = {Kuramochi, Eiichi and Notomi, Masaya and Mitsugi, Satoshi and Shinya, Akihiko and Tanabe, Takasumi and Watanabe, Toshifumi},
    title = {Ultrahigh-Q photonic crystal nanocavities realized by the local width modulation of a line defect},
    journal = {Applied Physics Letters},
    volume = {88},
    number = {4},
pages = {041112},
    year = {2006},
    month = {01},
    issn = {0003-6951},
    doi = {10.1063/1.2167801},
    url = {https://doi.org/10.1063/1.2167801},
    eprint = {https://pubs.aip.org/aip/apl/article-pdf/doi/10.1063/1.2167801/13140488/041112\_1\_online.pdf},
}

@Article{Shao2020,
author={Shao, Zeng-Kai
and Chen, Hua-Zhou
and Wang, Suo
and Mao, Xin-Rui
and Yang, Zhen-Qian
and Wang, Shao-Lei
and Wang, Xing-Xiang
and Hu, Xiao
and Ma, Ren-Min},
title={A high-performance topological bulk laser based on band-inversion-induced reflection},
journal={Nature Nanotechnology},
year={2020},
month={Jan},
day={01},
volume={15},
number={1},
pages = {67--72},
issn={1748-3395},
doi={10.1038/s41565-019-0584-x},
url={https://doi.org/10.1038/s41565-019-0584-x}
}

@Article{Sollner2015,
author={S{\"o}llner, Immo
and Mahmoodian, Sahand
and Hansen, Sofie Lindskov
and Midolo, Leonardo
and Javadi, Alisa
and Kir{\v{s}}ansk{\.{e}}, Gabija
and Pregnolato, Tommaso
and El-Ella, Haitham
and Lee, Eun Hye
and Song, Jin Dong
and Stobbe, S{\o}ren
and Lodahl, Peter},
title={Deterministic Photon--Emitter Coupling in Chiral Photonic Circuits},
journal={Nature Nanotechnology},
year={2015},
month={Sep},
day={01},
volume={10},
number={9},
pages = {775--778},
issn={1748-3395},
doi={10.1038/nnano.2015.159},
url={https://doi.org/10.1038/nnano.2015.159}
}

@article{PhysRevLett.62.2747,
  title = {Berry's phase for energy bands in solids},
  author = {Zak, J.},
  journal = {Phys. Rev. Lett.},
  volume = {62},
  issue = {23},
pages = {2747--2750},
  numpages = {0},
  year = {1989},
  month = {Jun},
  publisher = {American Physical Society},
  doi = {10.1103/PhysRevLett.62.2747},
  url = {https://link.aps.org/doi/10.1103/PhysRevLett.62.2747}
}

@article{PhysRevB.103.235110,
  title = {Topology of an anti-parity-time symmetric non-Hermitian Su-Schrieffer-Heeger model},
  author = {Wu, H. C. and Jin, L. and Song, Z.},
  journal = {Phys. Rev. B},
  volume = {103},
  issue = {23},
pages = {235110},
  numpages = {9},
  year = {2021},
  month = {Jun},
  publisher = {American Physical Society},
  doi = {10.1103/PhysRevB.103.235110},
  url = {https://link.aps.org/doi/10.1103/PhysRevB.103.235110}
}

@article{PhysRevB.94.195109,
  title = {Nonsymmorphic symmetry-required band crossings in topological semimetals},
  author = {Zhao, Y. X. and Schnyder, Andreas P.},
  journal = {Phys. Rev. B},
  volume = {94},
  issue = {19},
pages = {195109},
  numpages = {6},
  year = {2016},
  month = {Nov},
  publisher = {American Physical Society},
  doi = {10.1103/PhysRevB.94.195109},
  url = {https://link.aps.org/doi/10.1103/PhysRevB.94.195109}
}

@article{PhysRevB.106.064304,
  title = {Topological waves guided by a glide-reflection symmetric crystal interface},
  author = {Iglesias Mart\'{\i}nez, Julio Andr\'es and Laforge, Nicolas and Kadic, Muamer and Laude, Vincent},
  journal = {Phys. Rev. B},
  volume = {106},
  issue = {6},
pages = {064304},
  numpages = {6},
  year = {2022},
  month = {Aug},
  publisher = {American Physical Society},
  doi = {10.1103/PhysRevB.106.064304},
  url = {https://link.aps.org/doi/10.1103/PhysRevB.106.064304}
}

@Article{Plotnik2014,
author={Plotnik, Yonatan
and Rechtsman, Mikael C.
and Song, Daohong
and Heinrich, Matthias
and Zeuner, Julia M.
and Nolte, Stefan
and Lumer, Yaakov
and Malkova, Natalia
and Xu, Jingjun
and Szameit, Alexander
and Chen, Zhigang
and Segev, Mordechai},
title={Observation of unconventional edge states in `photonic graphene'},
journal={Nature Materials},
year={2014},
month={Jan},
day={01},
volume={13},
number={1},
pages = {57--62},
issn={1476-4660},
doi={10.1038/nmat3783},
url={https://doi.org/10.1038/nmat3783}
}

@Article{dai2023high,
author={Dai, Wei
and Yoda, Taiki
and Moritake, Yuto
and Ono, Masaaki
and Kuramochi, Eiichi
and Notomi, Masaya},
title={High transmission in 120-degree sharp bends of inversion-symmetric and inversion-asymmetric photonic crystal waveguides},
journal={Nature Communications},
year={2025},
month={Jan},
day={18},
volume={16},
number={1},
pages = {796},
issn={2041-1723},
doi={10.1038/s41467-025-56020-8},
url={https://doi.org/10.1038/s41467-025-56020-8}
}

@article{fukui2005chern,
  title={Chern Numbers in a Discretized {Brillouin} Zone: Efficient Method of Computing (Spin) {Hall} Conductances},
  author={Fukui, Takahiro and Hatsugai, Yasuhiro and Suzuki, Hiroshi},
  journal={Journal of the Physical Society of Japan},
  volume={74},
  number={6},
pages = {1674--1677},
  year={2005},
  publisher={The Physical Society of Japan}
}

@article{Mock:20,
author = {Adam Mock},
journal = {J. Opt. Soc. Am. B},
number = {1},
pages = {168--180},
publisher = {Optica Publishing Group},
title = {Symmetry-engineered waveguide dispersion in PT symmetric photonic crystal waveguides},
volume = {37},
month = {Jan},
year = {2020},
url = {https://opg.optica.org/josab/abstract.cfm?URI=josab-37-1-168},
doi = {10.1364/JOSAB.37.000168},
}

@article{PhysRevA.111.033513,
  title = {Exceptional-point restoration and fast-light edge states in photonic crystal waveguides with glide and time-reversal symmetries},
  author = {Uemura, Takahiro and Yoda, Taiki and Moritake, Yuto and Otsuka, Shutaro and Takata, Kenta and Notomi, Masaya},
  journal = {Phys. Rev. A},
  volume = {111},
  issue = {3},
pages = {033513},
  numpages = {15},
  year = {2025},
  month = {Mar},
  publisher = {American Physical Society},
  doi = {10.1103/PhysRevA.111.033513},
  url = {https://link.aps.org/doi/10.1103/PhysRevA.111.033513}
}

@article{PhysRevLett.106.106802,
  title = {Topological Crystalline Insulators},
  author = {Fu, Liang},
  journal = {Phys. Rev. Lett.},
  volume = {106},
  issue = {10},
pages = {106802},
  numpages = {4},
  year = {2011},
  month = {Mar},
  publisher = {American Physical Society},
  doi = {10.1103/PhysRevLett.106.106802},
  url = {https://link.aps.org/doi/10.1103/PhysRevLett.106.106802}
}

@Article{Huang2011,
author={Huang, Xueqin
and Lai, Yun
and Hang, Zhi Hong
and Zheng, Huihuo
and Chan, C. T.},
title={Dirac cones induced by accidental degeneracy in photonic crystals and zero-refractive-index materials},
journal={Nature Materials},
year={2011},
month={Aug},
day={01},
volume={10},
number={8},
pages = {582--586},
issn={1476-4660},
doi={10.1038/nmat3030},
url={https://doi.org/10.1038/nmat3030}
}

@article{Asano:17,
author = {Takashi Asano and Yoshiaki Ochi and Yasushi Takahashi and Katsuhiro Kishimoto and Susumu Noda},
journal = {Opt. Express},
number = {3},
pages = {1769--1777},
publisher = {Optica Publishing Group},
title = {Photonic crystal nanocavity with a Q factor exceeding eleven million},
volume = {25},
month = {Feb},
year = {2017},
url = {https://opg.optica.org/oe/abstract.cfm?URI=oe-25-3-1769},
doi = {10.1364/OE.25.001769},
}

@article{Takata:23,
author = {Kenta Takata and Eiichi Kuramochi and Akihiko Shinya and Masaya Notomi},
journal = {Opt. Express},
number = {7},
pages = {11864--11884},
publisher = {Optica Publishing Group},
title = {Improved design and experimental demonstration of ultrahigh-Q C6-symmetric H1 hexapole photonic crystal nanocavities},
volume = {31},
month = {Mar},
year = {2023},
url = {https://opg.optica.org/oe/abstract.cfm?URI=oe-31-7-11864},
doi = {10.1364/OE.485093},
}

@article{10.1063/5.0186703,
    author = {Zhan, Zi-Mei and Guo, Peng-Yu and Li, Wei and Wang, Hai-Xiao and Jiang, Jian-Hua},
    title = {Topological light guiding and trapping via shifted photonic crystal interfaces},
    journal = {Applied Physics Letters},
    volume = {123},
    number = {25},
pages = {251107},
    year = {2023},
    month = {12},
    issn = {0003-6951},
    doi = {10.1063/5.0186703},
    url = {https://doi.org/10.1063/5.0186703},
    eprint = {https://pubs.aip.org/aip/apl/article-pdf/doi/10.1063/5.0186703/18267302/251107_1_5.0186703.pdf},
}

@article{Hwang2024,
  author  = {Hwang, Min-Soo and Kim, Ha-Reem and Kim, Jungkil and Yang, Bohm-Jung and Kivshar, Yuri and Park, Hong-Gyu},
  title   = {Vortex nanolaser based on a photonic disclination cavity},
  journal = {Nature Photonics},
  year    = {2024},
  volume  = {18},
pages = {286--293},
  doi     = {10.1038/s41566-023-01338-2}
}

@article{Liu2021BulkDisclination,
  author  = {Liu, Yang and Leung, Shuwai and Li, Fei-Fei and Lin, Zhi-Kang and Tao, Xiufeng and Poo, Yin and Jiang, Jian-Hua},
  title   = {Bulk--disclination correspondence in topological crystalline insulators},
  journal = {Nature},
  year    = {2021},
  volume  = {589},
  number  = {7842},
pages = {381--385},
  doi     = {10.1038/s41586-020-03125-3},
  url     = {https://doi.org/10.1038/s41586-020-03125-3}
}

@article{Cui2025DisclinationNanocavities,
  author  = {Cui, Zihang and Guo, Wei and Hong, Xing and Tao, Jin and Liang, Guozhen and Zhu, Bofeng and Zeng, Yongquan},
  title   = {Photonic Disclination Nanocavities with Versatile Rotational Symmetries},
  journal = {Nano Letters},
  year    = {2025},
  volume  = {25},
  number  = {33},
pages = {12516--12523},
  doi     = {10.1021/acs.nanolett.5c02570},
  url     = {https://doi.org/10.1021/acs.nanolett.5c02570}
}

@article{Wang2025TopologicalCavities,
  author  = {Wang, Wenhao and Shen, Zhonglei and Tan, Yi Ji and Chen, Kaiji and Singh, Ranjan},
  title   = {On-chip topological edge state cavities},
  journal = {Light: Science \& Applications},
  year    = {2025},
  volume  = {14},
pages = {330},
  doi     = {10.1038/s41377-025-02017-3}
}

@article{Zhao2025SlowLightCavities,
  author  = {Zhao, Xin and Tan, Yi Ji and Wang, Wenhao and Chen, Kaiji and Shen, Zhonglei and Wang, Shixiong and Yi, Jianjia and Zhu, Lina and Singh, Ranjan},
  title   = {On-Chip Active High-{Q} Slow Light Topological Cavities},
  journal = {Laser \& Photonics Reviews},
  year    = {2025},
pages = {2402232},
  doi     = {10.1002/lpor.202402232}
}

@article{Hallacy2025,
  author = {Hallacy, L. and Martin, N. J. and Jalali Mehrabad, M. and Hallett, D. and Chen, X. and Dost, R. and Foster, A. and Brunswick, L. and Fenzl, A. and Clarke, E. and Patil, P. K. and Fox, A. M. and Skolnick, M. S. and Wilson, L. R.},
  title = {Nonlinear quantum optics at a topological interface enabled by defect engineering},
  journal = {npj Nanophotonics},
  volume = {2},
pages = {9},
  year = {2025},
  doi = {10.1038/s44310-025-00057-6},
  url = {https://doi.org/10.1038/s44310-025-00057-6}
}

@article{Li2025ValleyDefectCavity,
  author = {Li, Jiamu and Gao, Wenya and Gao, Yanyu and Zhang, Yanxia and Li, Xiaoxin and Jia, Qi and Shi, Bojian and Feng, Rui and Cao, Yongyin and Sun, Fangkui and Ding, Weiqiang},
  title = {Directly coupled low reflection loss valley topology defect cavity},
  journal = {Optics Letters},
  volume = {50},
  number = {16},
pages = {4958--4961},
  year = {2025},
  doi = {10.1364/OL.566394},
  url = {https://doi.org/10.1364/OL.566394}
}

@article{Rosiek2023Backscattering,
  author  = {Rosiek, Christian Anker and Arregui, Guillermo
             and Vladimirova, Anastasiia and Albrechtsen, Marcus
             and Vosoughi Lahijani, Babak
             and Christiansen, Rasmus Elleb{\ae}k and Stobbe, S{\o}ren},
  title   = {Observation of strong backscattering in valley-{H}all
             photonic topological interface modes},
  journal = {Nature Photonics},
  volume  = {17},
pages = {386--392},
  year    = {2023},
  doi     = {10.1038/s41566-023-01189-x}
}

@article{He2019ValleyRouting,
  author  = {He, Xin-Tao and Liang, En-Tao and Yuan, Jia-Jun and
             Qiu, Hao-Yang and Chen, Xiao-Dong and Zhao, Fu-Li and
             Dong, Jian-Wen},
  title   = {A silicon-on-insulator slab for topological valley transport},
  journal = {Nature Communications},
  volume  = {10},
pages = {872},
  year    = {2019},
  doi     = {10.1038/s41467-019-08881-z}
}

@article{Mehrabad2023ChiralFilter,
  author  = {Jalali Mehrabad, M. and Foster, A. P. and Martin, N. J. and
             Dost, R. and Clarke, E. and Patil, P. K. and
             Skolnick, M. S. and Wilson, L. R.},
  title   = {Chiral topological add--drop filter for integrated
             quantum photonic circuits},
  journal = {Optica},
  volume  = {10},
  number  = {3},
pages = {415--421},
  year    = {2023},
  doi     = {10.1364/OPTICA.481684}
}

@article{PhysRevLett.87.253902,
  title = {Extremely Large Group-Velocity Dispersion of Line-Defect Waveguides in Photonic Crystal Slabs},
  author = {Notomi, M. and Yamada, K. and Shinya, A. and Takahashi, J. and Takahashi, C. and Yokohama, I.},
  journal = {Phys. Rev. Lett.},
  volume = {87},
  issue = {25},
pages = {253902},
  numpages = {4},
  year = {2001},
  month = {Nov},
  publisher = {American Physical Society},
  doi = {10.1103/PhysRevLett.87.253902},
  url = {https://link.aps.org/doi/10.1103/PhysRevLett.87.253902}
}

@Article{Song2005,
author={Song, Bong-Shik
and Noda, Susumu
and Asano, Takashi
and Akahane, Yoshihiro},
title={Ultra-high-Q photonic double-heterostructure nanocavity},
journal={Nature Materials},
year={2005},
month={Mar},
day={01},
volume={4},
number={3},
pages = {207--210},
issn={1476-4660},
doi={10.1038/nmat1320},
url={https://doi.org/10.1038/nmat1320}
}
\end{document}